\pdfoutput=1

\RequirePackage{fix-cm}

\documentclass[twocolumn]{svjour3}

\smartqed

\usepackage{amsmath,amssymb}
\usepackage{graphicx}
\usepackage{txfonts}
\usepackage{natbib}
\usepackage{hyperref}

\journalname{Astrophysics and Space Science}

\begin{document}

\title{Revealing the basins of convergence in the planar equilateral restricted four-body problem}

\author{Euaggelos E. Zotos}

\institute{Department of Physics, School of Science, \\
Aristotle University of Thessaloniki, \\
GR-541 24, Thessaloniki, Greece\\
Corresponding author's email: {evzotos@physics.auth.gr}}

\date{Received: 29 September 2016 / Accepted: 7 November 2016 / Published online: 1 December 2016}

\titlerunning{Revealing the basins of convergence in the planar equilateral restricted four-body problem}

\authorrunning{Euaggelos E. Zotos}

\maketitle

\begin{abstract}

The planar equilateral restricted four-body problem where two of the primaries have equal masses is used in order to determine the Newton-Raphson basins of convergence associated with the equilibrium points. The parametric variation of the position of the libration points is monitored when the value of the mass parameter $m_3$ varies in predefined intervals. The regions on the configuration $(x,y)$ plane occupied by the basins of attraction are revealed using the multivariate version of the Newton-Raphson iterative scheme. The correlations between the attracting domains of the equilibrium points and the corresponding number of iterations needed for obtaining the desired accuracy are also illustrated. We perform a thorough and systematic numerical investigation by demonstrating how the dynamical parameter $m_3$ influences the shape, the geometry and the degree of fractality of the converging regions. Our numerical outcomes strongly indicate that the mass parameter is indeed one of the most influential factors in this dynamical system.


\end{abstract}

\section{Introduction}
\label{intro}

The topic of dynamical systems of few-bodies has always been one of the most fascinating fields in celestial mechanics and dynamical astronomy. Especially these days, with the detection of more than 3500 extra-solar planetary systems (see \url{http://exoplanets.eu}, update: November 3, 2016), the few-problem problem strongly attracts the scientific interest.

There is no doubt that one of the most well investigated versions of the few-body problem is the circular or elliptic restricted (or not) three-body problem \citep{S67}. In the same vein, the planar restricted four-body problem describes the motion of a test particle with infinitesimal mass (with respect to the masses of the primaries) moving inside the gravitational field of three primary bodies. There are two main configurations regarding the position of the three primary bodies: (i) the Eulerian configuration, where all three primaries lie on the same axis and (ii) the Lagrangian or triangular configuration, where the three primaries always lie at the vertices of an equilateral triangle. For the latter configuration we have the case of the planar equilateral restricted four-body problem (PERFBP). Usually, for the corresponding configurations we use the term ``central configurations" due to the fact that the accelerations of the three primary bodies are proportional to the corresponding radius-vectors, while they are directed toward the common center of gravity \citep{M90}.

Over the years, the four-body problem has been used, by many researchers, for several practical applications, such as describing real celestial systems. For example: \citet{VHW86,KSM10} for the epsilon Aurigae system, \citet{RG06,MLJ08} for the Sun-Jupiter-Saturn system, \citet{SSD09a} for a system of a star, two massive planets and a massless Trojan, \citet{SSD09b} and references therein for a system of a star, a brown dwarf, a gas giant and a massless Trojan, \citet{CB10} and references therein and \citet{BP13} for the Sun-Jupiter-Trojan Asteroid, Spacecraft system, \citet{SGJM95,J00,dAP05,MdS07} for the Sun-Earth-Moon system, where the fourth body can be a space vehicle.

A very interesting topic of the four-body problem is the location of the periodic orbits \citep[see e.g.,][]{SPB08,BP11b,BGD13,AB15}. Similarly, the determination of the position and the stability of the equilibrium points of the four-body problem is another issue of great importance \citep[see e.g.,][]{S78,H80,M81,L06,P07,BP11a,PP13,ASS15,SV15,SV16}.

In dynamical systems knowing the basins of attraction associated with the equilibrium points is very important since this knowledge reveals some of the most inartistic properties of the system. For obtaining the basins of convergence we use an iterative scheme and we perform a scan of the configuration $(x,y)$ plane in order to determine from which of the equilibrium points (attractors) each initial condition is attracted by. The attracting domains in several types of dynamical systems have been numerically investigated. The Newton-Raphson iterative method was used in \citet{D10} to explore the basins of attraction in the Hill's problem with oblateness and radiation pressure, while in \citet{Z16a} the multivariate version of the same iterative scheme has been used to unveil the basins of convergence in the restricted three-body problem with oblateness and radiation pressure. Furthermore, the Newton-Raphson converging domains for the photogravitational Copenhagen problem \citep[see e.g.,][]{K08}, the electromagnetic Copenhagen problem \citep[see e.g.,][]{KGK12}, the four-body problem \citep[see e.g.,][]{BP11a,KK14}, the ring problem of $N + 1$ bodies \citep[see e.g.,][]{CK07,GKK09}, or even the restricted 2+2 body problem \citep[see e.g.,][]{CK13} have been studied.

In this paper we shall work as in \citet{Z16a}, thus following the same numerical techniques and methodology, and we will try to reveal the Newton-Raphson basins of attraction on the configuration $(x,y)$ plane for special case of the PERFBP where two of the three primary bodies have equal masses.

The present paper is organized as follows: In Section \ref{mod} we present the basic properties of the considered mathematical model. In section \ref{lgevol} the parametric evolution of the position of the equilibrium points is investigated as the value of the mass parameter $m_3$ varies in predefined intervals. In the following Section, we conduct a thorough and systematic numerical exploration by revealing the Newton-Raphson basins of attraction of the PERFBP with two equal masses and how they are affected by the value of the mass parameter. Our paper ends with Section \ref{conc}, where the main conclusions of this work are presented.

\section{Presentation of the mathematical model}
\label{mod}

We consider a system of units where the units of length, mass and time are taken in such a way so that the sum of the masses $(m_1 + m_2 + m_3)$, the distance between the primaries $(R)$ and of course the angular velocity $(\omega)$ to be equal to unity. Consequently, the gravitational constant is $G = 1$.

\begin{figure}[!t]
\centering
\resizebox{\hsize}{!}{\includegraphics{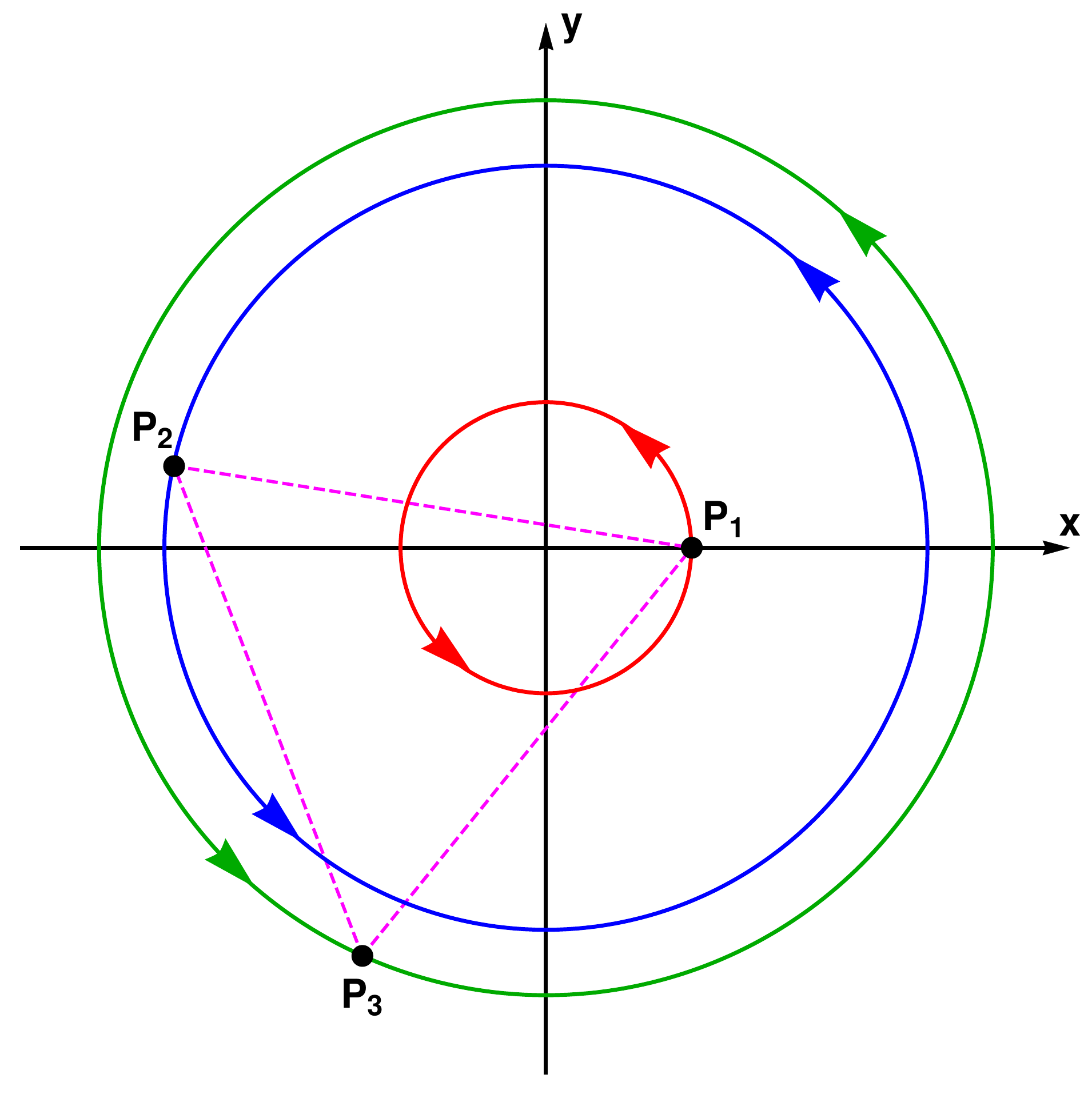}}
\caption{Equilateral triangular (Lagrangian) configuration of the three primary bodies, moving in circular orbits around their common center of gravity. In this case, $m_2 = 0.25$, $m_3 = 0.05$, while $m_1 = 1 - m_2 - m_3 = 0.7$.}
\label{erfbp}
\end{figure}

Without the loss of generality, we may assume that one of the primary bodies is located on the positive $x$ axis, at the origin of time. For describing the motion of the four-body system, we use axes rotating with uniform angular velocity. Furthermore, the three primaries move always on the $(x,y)$ plane, while their mutual distances remain constant with respect to the time (see Fig. \ref{erfbp}). In this study, we consider only circular orbits of the primaries around their common center of gravity.

The effective potential function in a synodic system of coordinates is defined as
\begin{equation}
\Omega(x,y) = \sum\limits_{i=1}^3 \frac{m_i}{r_i} + \frac{1}{2}\left(x^2 + y^2\right),
\label{pot}
\end{equation}
where
\begin{equation}
r_i = \sqrt{\left(x - x_i\right)^2 + \left(y - y_i\right)^2}, \ \ i = 1,2,3,
\label{dist}
\end{equation}
are the distances to the respective primaries.

Under the assumption that the three primaries move always on the $(x,y)$ plane and the axes are chosen in such a way so the $m_1$ is on the $x$ axis the coordinates of the primaries are given by the following relations \citep[see also][]{BP11b}
\begin{align}
&x_1 = \frac{|K|M}{K}, \nonumber\\
&y_1 = 0, \nonumber\\
&x_2 = - \frac{|K|\left[\left(m_2 - m_3\right)m_3 + m_1\left(2m_2 + m_3\right)\right]}{2KM}, \nonumber\\
&y_2 = \frac{\sqrt{3}}{2} \frac{m_3}{m_2^{3/2}}\frac{\sqrt{m_2^3}}{M}, \nonumber\\
&x_3 = - \frac{|K|}{2\sqrt{M}}, \nonumber\\
&y_3 = - \frac{\sqrt{3}}{2} \frac{1}{m_2^{1/2}}\frac{\sqrt{m_2^3}}{M},
\label{cents}
\end{align}
where
\begin{align}
&|K| = m_2\left(m_3 - m_2\right) + m_1\left(m_2 + 2m_3\right), \nonumber\\
&M = \sqrt{m_2^2 + m_2m_3 + m_3^2}.
\end{align}

The above-mentioned relations (\ref{cents}) apply for the general case where $m_1 \neq m_2 \neq m_3$. In the special case where $m_2 = m_3 < 1/2$ and of course $m_1 = 1 - 2m_3$ the coordinates of the centers of the three primaries are: $P_1(x_1, y_1) = (m_3\sqrt{3}, 0)$, $P_2(x_2, y_2) = (\left(2m_3 - 1\right)\sqrt{3}/2, 1/2)$, and $P_3(x_3, y_3) = (x_2, -y_2)$.

The equations describing the motion of an infinitesimal mass (test particle) in the usual dimensionless rectangular rotating coordinate system read \citep{M00}
\begin{align}
&\Omega_x(x,y) = \frac{\partial \Omega}{\partial x} = \ddot{x} - 2\dot{y} = x - \sum\limits_{i=1}^3 \frac{m_i\left(x - x_i\right)}{r_i^3}, \nonumber\\
&\Omega_y(x,y) = \frac{\partial \Omega}{\partial y} = \ddot{y} + 2\dot{x} = y - \sum\limits_{i=1}^3 \frac{m_i\left(y - y_i\right)}{r_i^3},
\label{eqmot}
\end{align}
where dots denote the time derivatives.

In the same vein, the second order derivatives (which will be needed for the multivariate Newton-Raphson iterative scheme) are written as
\begin{align}
&\Omega_{xx}(x,y) = \frac{\partial^2 \Omega}{\partial x^2} = 1 + \sum\limits_{i=1}^3 \frac{m_i\left[2\left(x - x_i\right)^2 - \left(y - y_i\right)^2\right]}{r_i^5}, \nonumber\\
&\Omega_{xy}(x,y) = \frac{\partial^2 \Omega}{\partial x \partial y} = 3\sum\limits_{i=1}^3 \frac{m_i\left(x - x_i\right)\left(y - y_i\right)}{r_i^5}, \nonumber\\
&\Omega_{yx}(x,y) = \frac{\partial^2 \Omega}{\partial y \partial x} = \Omega_{xy}(x,y), \nonumber\\
&\Omega_{yy}(x,y) = \frac{\partial^2 \Omega}{\partial y^2} = 1 - \sum\limits_{i=1}^3 \frac{m_i\left[\left(x - x_i\right)^2 - 2\left(y - y_i\right)^2\right]}{r_i^5}.
\label{vareq}
\end{align}

The system of differential equations (\ref{eqmot}) admits the integral of the total orbital energy (also known as the Jacobi integral of motion)
\begin{equation}
J(x,y,\dot{x},\dot{y}) = 2\Omega(x,y) - \left(\dot{x}^2 + \dot{y}^2 \right) = C,
\label{ham}
\end{equation}
where $\dot{x}$ and $\dot{y}$ are the velocities, while $C$ is the Jacobi constant which is conserved.

\section{Parametric variation of the equilibrium points}
\label{lgevol}

It is well known that the necessary and sufficient conditions for the existence of every equilibrium point are
\begin{equation}
\dot{x} = \dot{y} = \ddot{x} = \ddot{y} = 0.
\label{lps0}
\end{equation}
Therefore, the coordinates of the positions of all the coplanar equilibrium points of the PERFBP can be numerically derived by solving the following system of partial differential equations
\begin{equation}
\begin{cases}
\Omega_x(x,y) = 0 \\
\Omega_y(x,y) = 0
\end{cases}.
\label{lps}
\end{equation}

\begin{figure}[!t]
\centering
\resizebox{0.77\hsize}{!}{\includegraphics{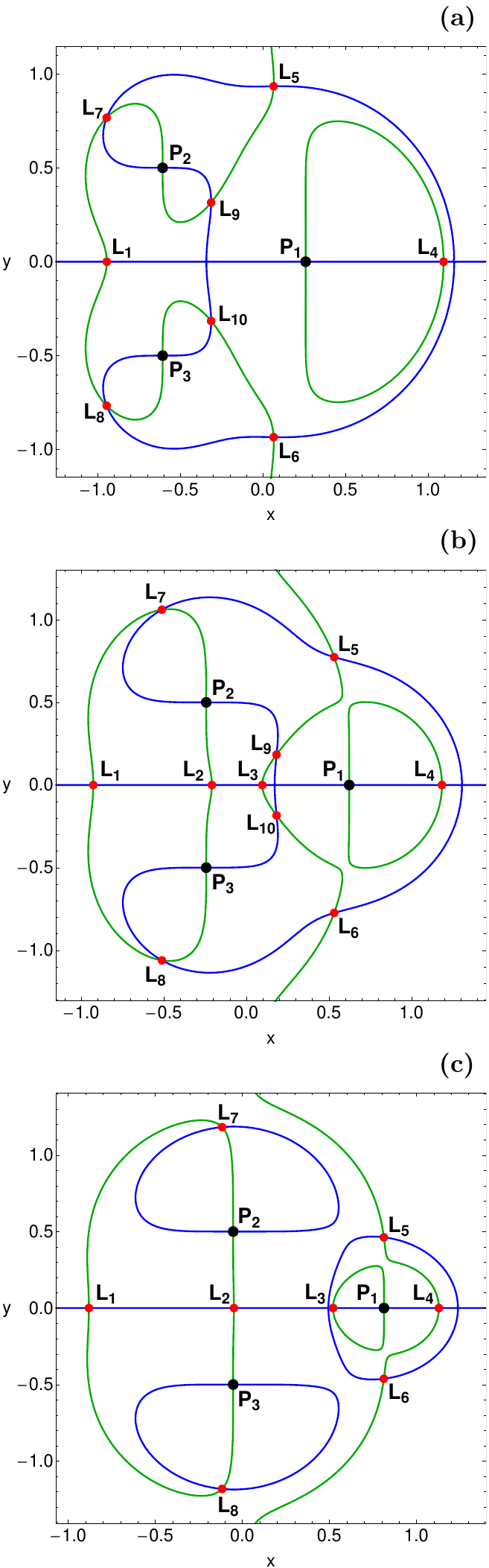}}
\caption{Locations of the positions (red dots) of the equilibrium points $(L_i, \ i=1,10)$ through the intersections of $\Omega_x = 0$ (green) and $\Omega_y = 0$ (blue), when (a-upper panel): $m_3 = 0.15$, (b-middle panel): $m_3 = 0.36$, and (c-lower panel): $m_3 = 0.47$. The black dots denote the centers $(P_i, \ i=1,3)$ of the three primary bodies.}
\label{conts}
\end{figure}

At this point, we would like to emphasize that for the rest of the paper we shall deal only with the case where two of the three primaries have equal masses $(m_2 = m_3)$. When $y_0 = 0$ the second of equations (\ref{eqmot}) is fully satisfied because there are only terms of $x_2$, $y_2$, $x_3$, and $y_3$ which cancel each other due to the symmetry of the system. Therefore, for the PERFBP with two equal masses collinear equilibrium points exist for every possible value of the mass parameter $m_3$. The coordinates $(x_0,0)$ of the collinear equilibrium points are obtained from the first of equations (\ref{eqmot}) for $y_0 = 0$.

In \citet{BP11a} it was shown that the total number of the equilibrium points in the PERFBP with two equal masses in not constant but it depends on the value of the mass parameter $m_3$. In particular
\begin{itemize}
  \item When $m_3 \in (0,0.2882761]$ there are eight equilibrium points: two collinear and six non-collinear points (see panel (a) of Fig. \ref{conts}).
  \item When $m_3 \in [0.2882762,0.4402]$ there are ten equilibrium points: four collinear and six non-collinear points (see panel (b) of Fig. \ref{conts}).
  \item When $m_3 \in [0.4403,0.5)$ there are eight equilibrium points: four collinear and four non-collinear points (see panel (c) of Fig. \ref{conts}).
\end{itemize}

In Fig. \ref{conts} we see how the intersections of equations $\Omega_x = 0, \ \Omega_y = 0$ define on the configuration $(x,y)$ plane the positions of the equilibrium points when (a): $m_3 = 0.15$, (b): $m_3 = 0.36$, and (c): $m_3 = 0.47$.

\begin{figure}
\includegraphics[width=\hsize]{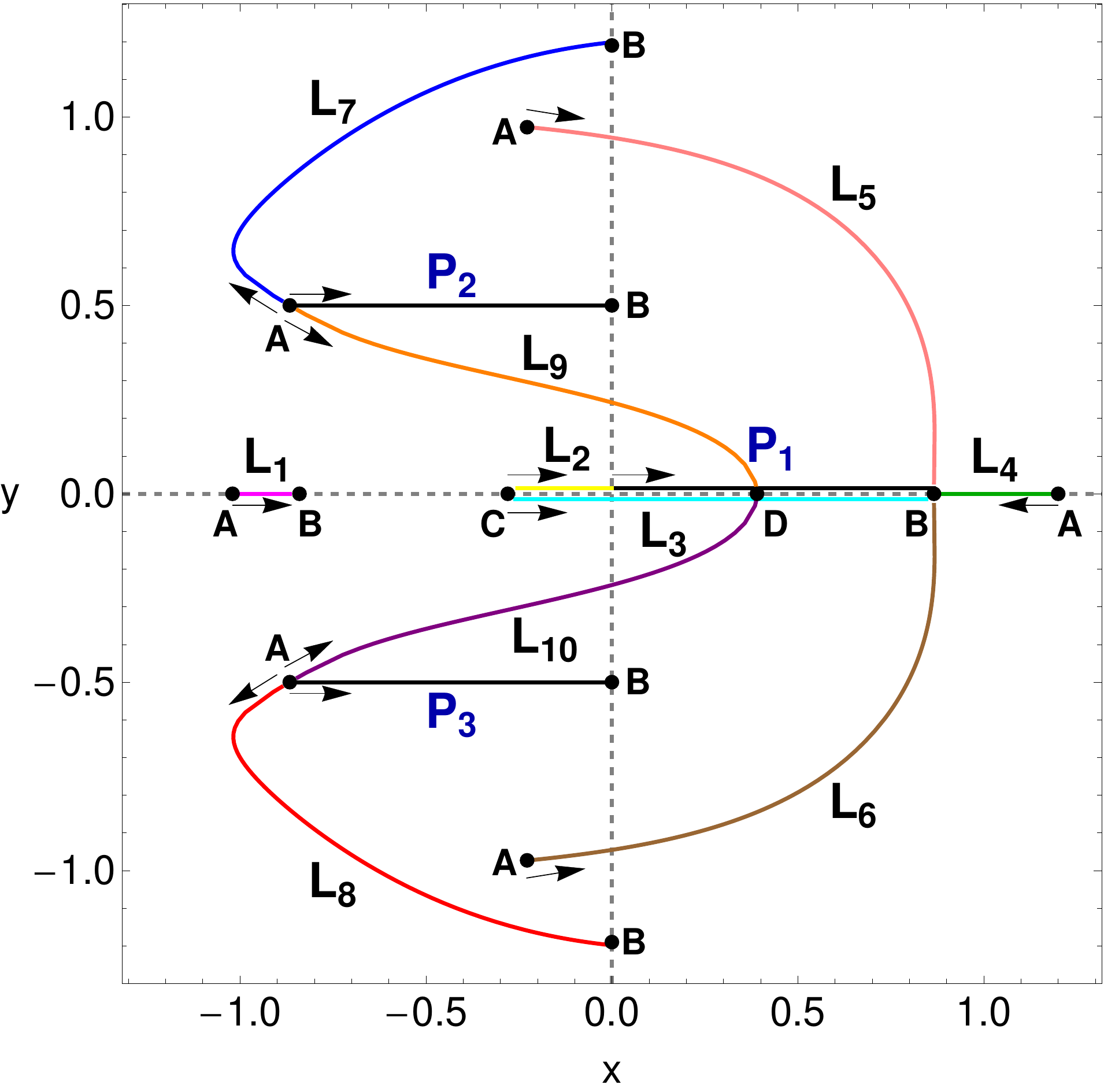}
\caption{The space-evolution of the equilibrium points in the PERFBP with two equal masses, when $m_3 \in (0, 1/2)$. The arrows indicate the direction of the movement of both the equilibrium points and the centers of the primary bodies, as the mass parameter $m_3$ increases. The black dots correspond to critical values of the mass parameter $m_3$. The meaning of the capital letters, regarding the critical and asymptotic values of $m_3$, is the following: A: $m_3 \to 0$, B: $m_3 \to 0.5$, C: $m_3 \to 0.2882762$, and D: $m_3 \to 0.4403$.}
\label{lgs}
\end{figure}

In this investigation we shall reveal how the mass parameter $m_3$ influences the positions of the equilibrium points, when it varies in the interval $m_3 \in (0, 1/2)$. Our results are illustrated in Fig. \ref{lgs}, where we present the space-evolution of all the equilibrium points on the configuration $(x,y)$ plane. One may observe that as value of the mass parameter $m_3$ tends to zero the equilibrium points $L_7$ and $L_9$, as well as $L_8$ and $L_{10}$ move towards the centers $P_2$ and $P_3$, respectively. It is seen that the libration points $L_2$ and $L_3$ emerge only when $m_3 \geq 0.2882762$ ($L_2$ and $L_3$ completely coincide for $m_3$ = 0.2882762). Furthermore, at the special case where $m_3 \to 0.4402$ it was found that the equilibrium points $L_9$ and $L_{10}$ collide with each other on the $x$ axis and they disappear for higher values of the mass parameter. Furthermore, it it interesting to note that as the value of the mass parameter tends to 1/2 the libration points $L_3$, $L_4$, $L_5$, and $L_6$ tend to collide with each other on the $x$-axis. In fact, when $m_3 = 1/2$ only five equilibrium points exist since the PERFBP degenerates to the restricted three-body problem with equal masses (also known as the Copenhagen problem). In addition, for $m_3 \to 1/2$ the equilibrium points $L_7$ and $L_8$ as well as the centers of the primary bodies 2 and 3 tend to the vertical $y$ axis.

The center of the primary 1 $(P_1)$ moves always on the $x$ axis in the interval $(0, \sqrt{3}/2)$, while the centers $P_2$ and $P_3$ of the primaries with equal masses move on the parallel lines $y = 1/2$ and $y = -1/2$, respectively in the interval $(-\sqrt{3}/2,0)$, when the value of the mass parameter varies in the interval $(0,1/2)$. In the PERFBP with two equal masses $(m_2 = m_3)$ the $x$ axis $(y = 0)$ is the only axis of symmetry (observe in Fig. \ref{lgs} the symmetry of the parametric evolution of all points with respect to the $x$ axis).

The stability of all the equilibrium points of the PERFBP with two equal masses has been numerically investigated in \citet{BP11a} (see Table 1).

\section{The basins of attraction}
\label{bas}

There is no doubt that the most famous numerical method for solving systems of equations is the Newton-Raphson method. This method is also applicable to systems of multivariate functions $f({\bf{x}}) = 0$, through the iterative scheme
\begin{equation}
{\bf{x}}_{n+1} = {\bf{x}}_{n} - J^{-1}f({\bf{x}}_{n}),
\label{sch}
\end{equation}
where $J^{-1}$ is the inverse Jacobian matrix of the system of differential equations $f({\bf{x_n}})$, where in our case it is described in Eqs. (\ref{lps}).

With trivial matrix calculations \citep[see e.g., Appendix in][]{Z16b} we can obtain the following iterative formulae for each coordinate
\begin{eqnarray}
x_{n+1} &=& x_n - \left( \frac{\Omega_x \Omega_{yy} - \Omega_y \Omega_{xy}}{\Omega_{yy} \Omega_{xx} - \Omega^2_{xy}} \right)_{(x_n,y_n)}, \nonumber\\
y_{n+1} &=& y_n + \left( \frac{\Omega_x \Omega_{yx} - \Omega_y \Omega_{xx}}{\Omega_{yy} \Omega_{xx} - \Omega^2_{xy}} \right)_{(x_n,y_n)},
\label{nrm}
\end{eqnarray}
where $x_n$, $y_n$ are the values of the $x$ and $y$ coordinates at the $n$-th step of the iterative process, while the subscripts denote the corresponding partial derivatives of first and second order of the effective potential function $\Omega(x,y)$.

\begin{figure}[!t]
\includegraphics[width=\hsize]{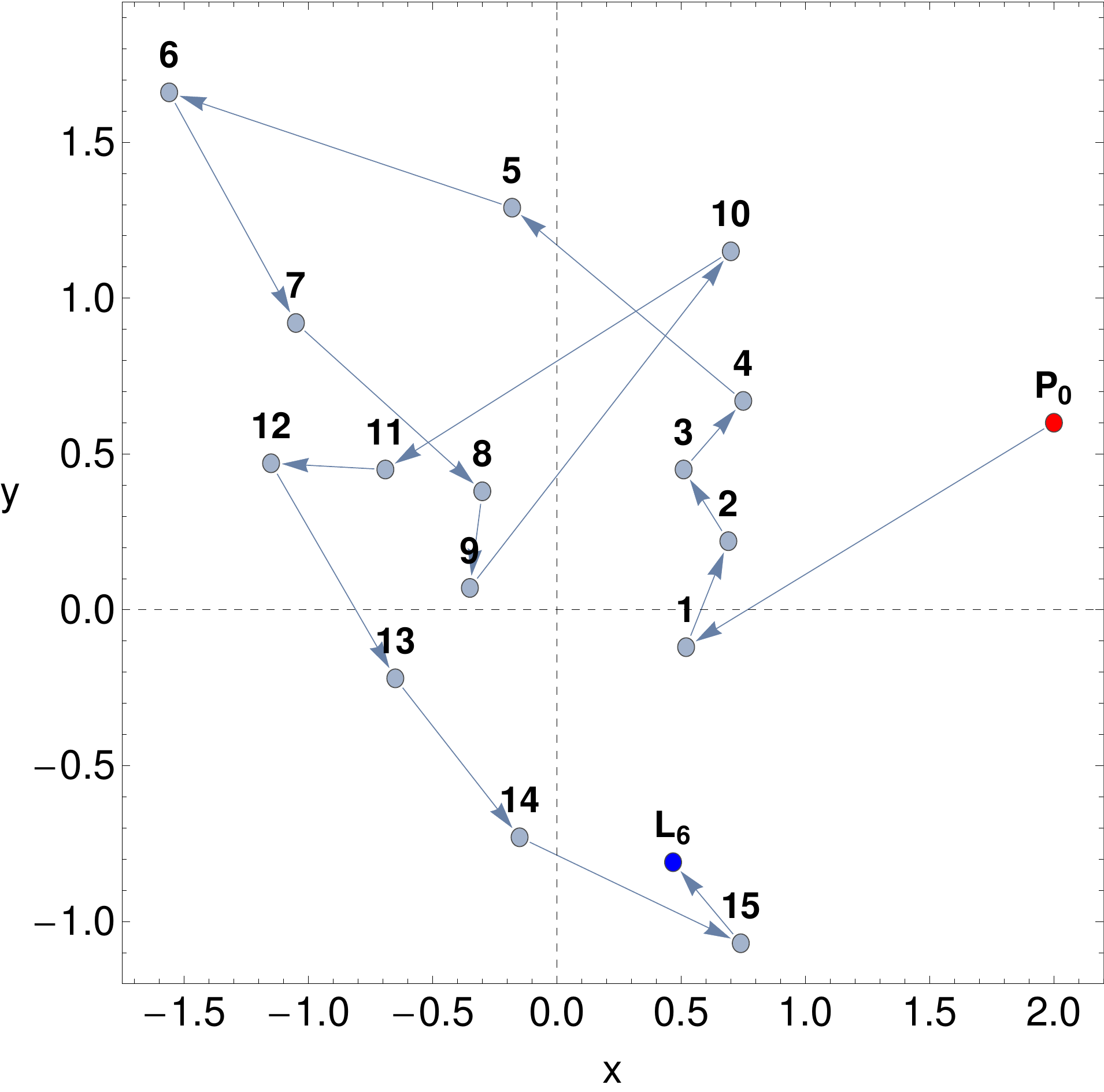}
\caption{A characteristic example of the consecutive steps that are followed by the Newton-Raphson iterator and the corresponding crooked path-line that leads to an equilibrium point $(L_6)$, when $m_3 = 1/3$. The red dot indicates the starting point $P_0$ with $(x_0, y_0) = (2.9, 0.6)$, while the blue dot indicates the equilibrium point to which the method converged to. For this particular set of initial conditions the Newton-Raphson method converges after 16 iterations to $L_6$ with accuracy of six decimal digits, while only three more iterations are required for obtaining the desired accuracy of $10^{-15}$.}
\label{nr}
\end{figure}

The Newton-Raphson algorithm works as follows: an initial condition $(x_0,y_0)$ on the configuration plane activates the code and the iterative process continues until one of the equilibrium points of the system is reached, with some predefined accuracy. In most of the cases the successive approximation points create a crooked path line (see Fig. \ref{nr}). The initial condition may or may not converge to one of the libration points which act as attractors. If the crooked path leads to one of the equilibrium point then the iterative method converges for the particular initial condition. A Newton-Raphson basin of attraction\footnote{It should be clarified and clearly emphasized that the Newton-Raphson basins of convergence should not be mistaken, by no means, with the classical basins of attraction which exist in dissipative systems. The difference between the Newton-Raphson basins of convergence and the basins of attraction in dissipative systems is huge. This is true because the attraction in the first case is just a numerical artifact of the Newton-Raphson iterative method, while in dissipative systems the attraction is a real dynamical phenomenon, observed through the numerical integration of the initial conditions.} or convergence (also known as attracting region or domain) is composed of all the initial conditions that lead to a specific attractor (equilibrium point).

One may claim that knowing the basins of attraction of a dynamical system is an issue of paramount importance because these attracting regions may reflect some of the most important qualitative properties of the system in question. This can be justified by taking into account the fact that the derivatives of both first and second order of the effective potential function $\Omega(x,y)$ are included in the iterative formulae (\ref{nrm}).

\begin{figure*}[!t]
\centering
\resizebox{\hsize}{!}{\includegraphics{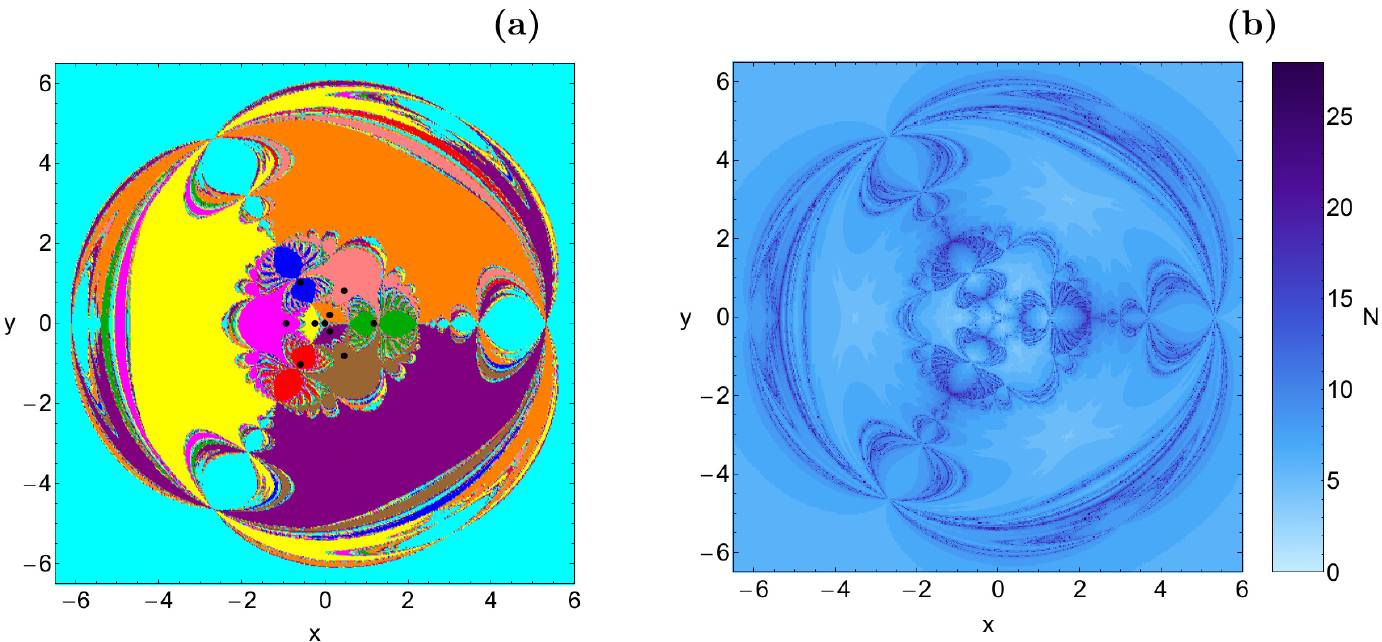}}
\caption{(a-left): The Newton-Raphson basins of attraction on the configuration $(x,y)$ plane for the case of three equal masses $(m_3 = 1/3)$. The positions of the ten equilibrium points are indicated by black dots. The color code denoting the ten attractors (equilibrium points) is as follows: $L_1$ (magenta); $L_2$ (yellow); $L_3$ (cyan); $L_4$ (green); $L_5$ (pink); $L_6$ (brown); $L_7$ (blue); $L_8$ (red); $L_9$ (orange); $L_{10}$ (purple); non-converging points (white). (b-right): The distribution of the corresponding number $(N)$ of required iterations for obtaining the Newton-Raphson basins of attraction shown in panel (a).}
\label{sm}
\end{figure*}

For revealing the structures of the basins of attraction on the configuration $(x,y)$ plane we define a dense uniform grid\footnote{Needless to say the the initial conditions corresponding to the three centers $(P_1, P_2, P_3)$ of the primaries are excluded from the grid because for these values the distances $r_i$, $i = 1,2,3$ to the primaries are zero and therefore several terms of the formulae (\ref{nrm}) become singular.} of $1024 \times 1024$ initial conditions (nodes), which will be used as the initial values of the numerical algorithm. The iterative procedure begins and stops only when an accuracy of $10^{-15}$ regarding the position of the attractors has been achieved. A double scanning of the configuration plane is performed in order to classify all the available initial conditions that lead to a specific equilibrium point (or attractor). While classifying the initial conditions we also record the number $N$ of required iterations in order to obtain the aforementioned accuracy. It is evident that there is a strong correlation between the required number of iterations and the desired accuracy; the better the accuracy the higher the required iterations. In this study we set the maximum number of iterations $N_{\rm max}$ to be equal to 500.

In panel (a) of Fig. \ref{sm} we present the Newton-Raphson basins of attraction when $m_3 = 1/3$, which means that all three primaries have equal masses. For each basin of convergence we use different color, while the positions of all the attractors (equilibrium points) are pinpointed by small black dots. All non-converging points are shown in white. We observe that in the case of three equal masses there are two additional axes of symmetry, $y = \sqrt{3}x$ and $y = - \sqrt{3}x$, along with the $y = 0$, which exists in the case of two equal masses. In panel (b) of the same figure the distribution of the corresponding number $(N)$ of iterations required for obtaining the desired accuracy is given using tones of blue. Looking the color-coded plot in Fig. \ref{sm}a we may say that the shape of the basins of convergence corresponding to equilibrium points $L_4$, $L_7$, and $L_8$ look like exotic bugs with many legs and many antennas, while the shape of the basins of attraction corresponding to all other libration points look like butterfly wings.

In the following we shall try to determine how the mass parameter $m_3$ influences the structure of the Newton-Raphson basins of attraction, considering three cases regarding the total number and also the type (collinear and non-collinear) of the equilibrium points.

For the classification of the initial conditions on the $(x,y)$ plane we will use modern color-coded diagrams. In these diagrams, each pixel is assigned a specific color according to the particular attractor (equilibrium point). The size of the two-dimensional grids, or in other words the minimum and the maximum values of $x$ and $y$, are chosen differently in each case so as to have a complete view of the basin structures created by the attractors.

\subsection{Case I: Two collinear points and six non-collinear points}
\label{ss1}

\begin{figure*}[!t]
\centering
\resizebox{\hsize}{!}{\includegraphics{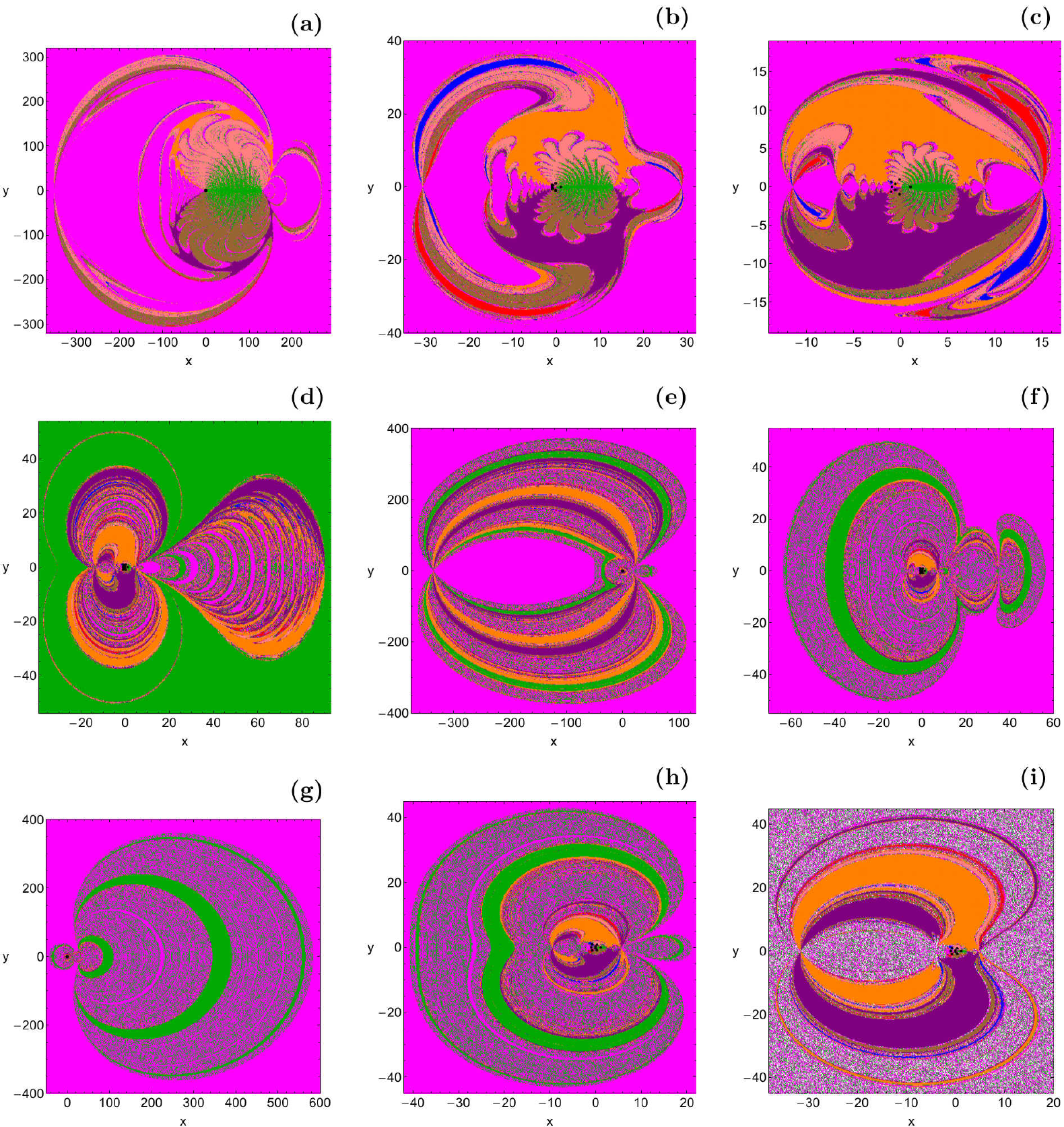}}
\caption{The Newton-Raphson basins of attraction on the configuration $(x,y)$ plane for the first case, where two collinear and six non-collinear equilibrium points are present. (a): $m_3 = 0.0001$; (b): $m_3 = 0.01$; (c): $m_3 = 0.05$; (d): $m_3 = 0.22$; (e): $m_3 = 0.25$; (f): $m_3 = 0.258$; (g): $m_3 = 0.26$; (h): $m_3 = 0.275$; (i): $m_3 = 0.2882761$. The positions of the eight equilibrium points are indicated by black dots. The color code, denoting the eight attractors and the non-converging points, is as in Fig. \ref{sm}.}
\label{pr1}
\end{figure*}

\begin{figure*}[!t]
\centering
\resizebox{\hsize}{!}{\includegraphics{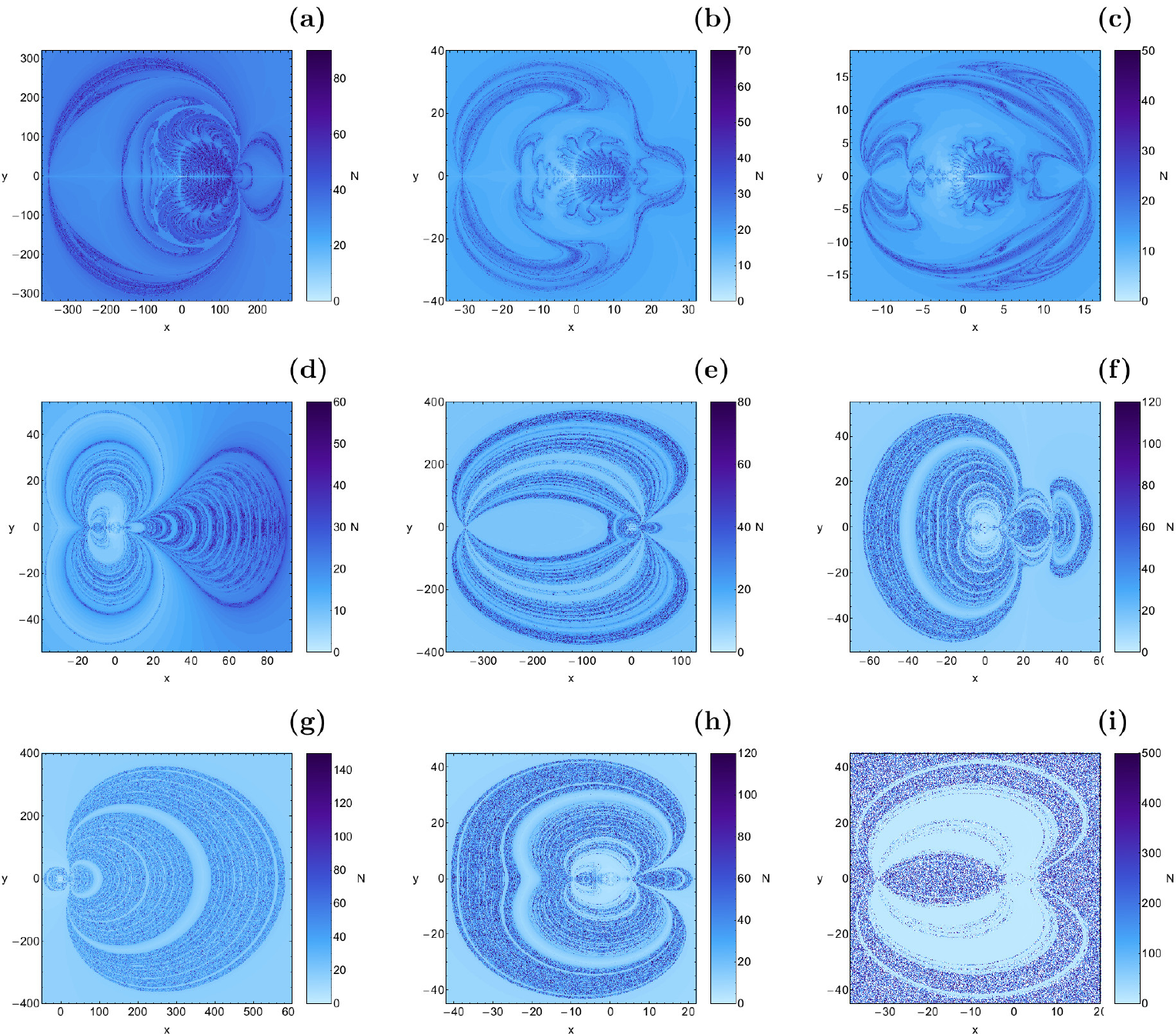}}
\caption{The distribution of the corresponding number $(N)$ of required iterations for obtaining the Newton-Raphson basins of attraction shown in Fig. \ref{pr1}(a-i). The non-converging points are shown in white.}
\label{pr1n}
\end{figure*}

\begin{figure*}[!t]
\centering
\resizebox{\hsize}{!}{\includegraphics{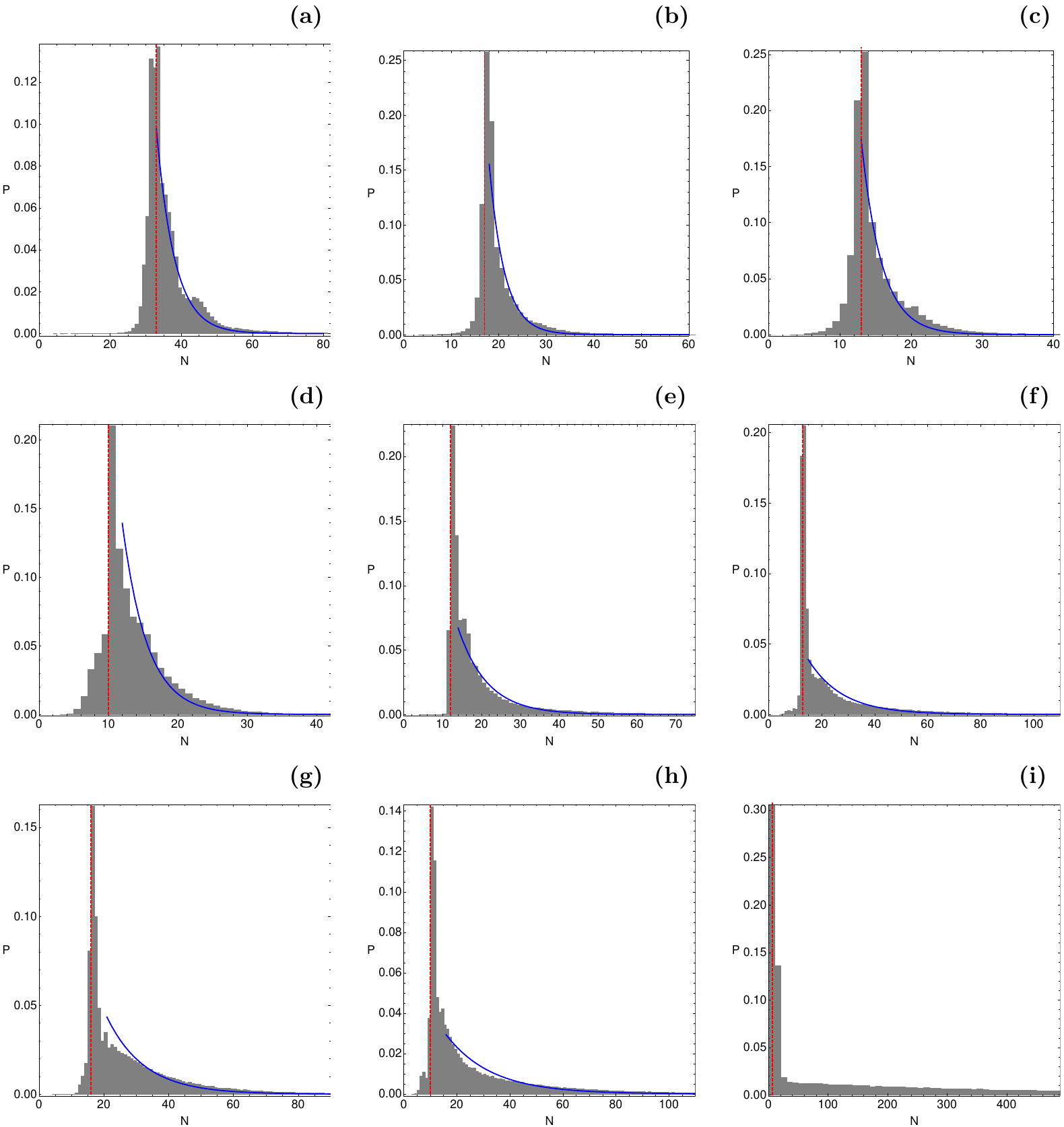}}
\caption{The corresponding probability distribution of required iterations for obtaining the Newton-Raphson basins of attraction shown in Fig. \ref{pr1}(a-i). The vertical dashed red line indicates, in each case, the most probable number $(N^{*})$ of iterations. The blue line is the best fit for the right-hand side $(N > N^{*})$ of the histograms, using a Laplace probability distribution function.}
\label{pr1p}
\end{figure*}

\begin{figure*}[!t]
\centering
\resizebox{\hsize}{!}{\includegraphics{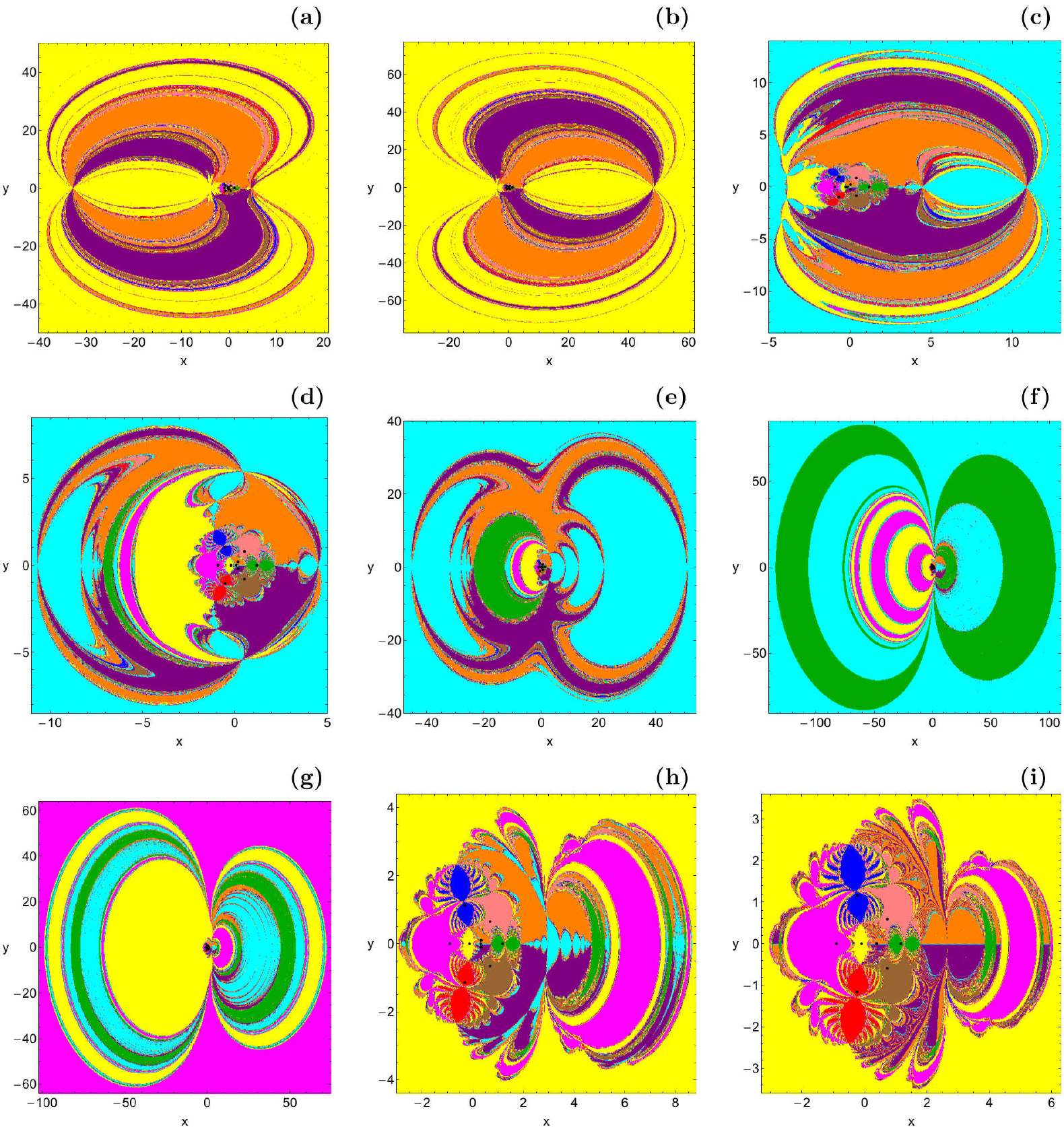}}
\caption{The Newton-Raphson basins of attraction on the configuration $(x,y)$ plane for the second case, where four collinear and six non-collinear equilibrium points are present. (a): $m_3 = 0.2882762$; (b): $m_3 = 0.29$; (c): $m_3 = 0.30$; (d): $m_3 = 0.35$; (e): $m_3 = 0.37$; (f): $m_3 = 0.38$; (g): $m_3 = 0.40$; (h): $m_3 = 0.42$; (i): $m_3 = 0.4402$. The positions of the ten equilibrium points are indicated by black dots. The color code, denoting the ten attractors and the non-converging points, is as in Fig. \ref{sm}.}
\label{pr2}
\end{figure*}

\begin{figure*}[!t]
\centering
\resizebox{\hsize}{!}{\includegraphics{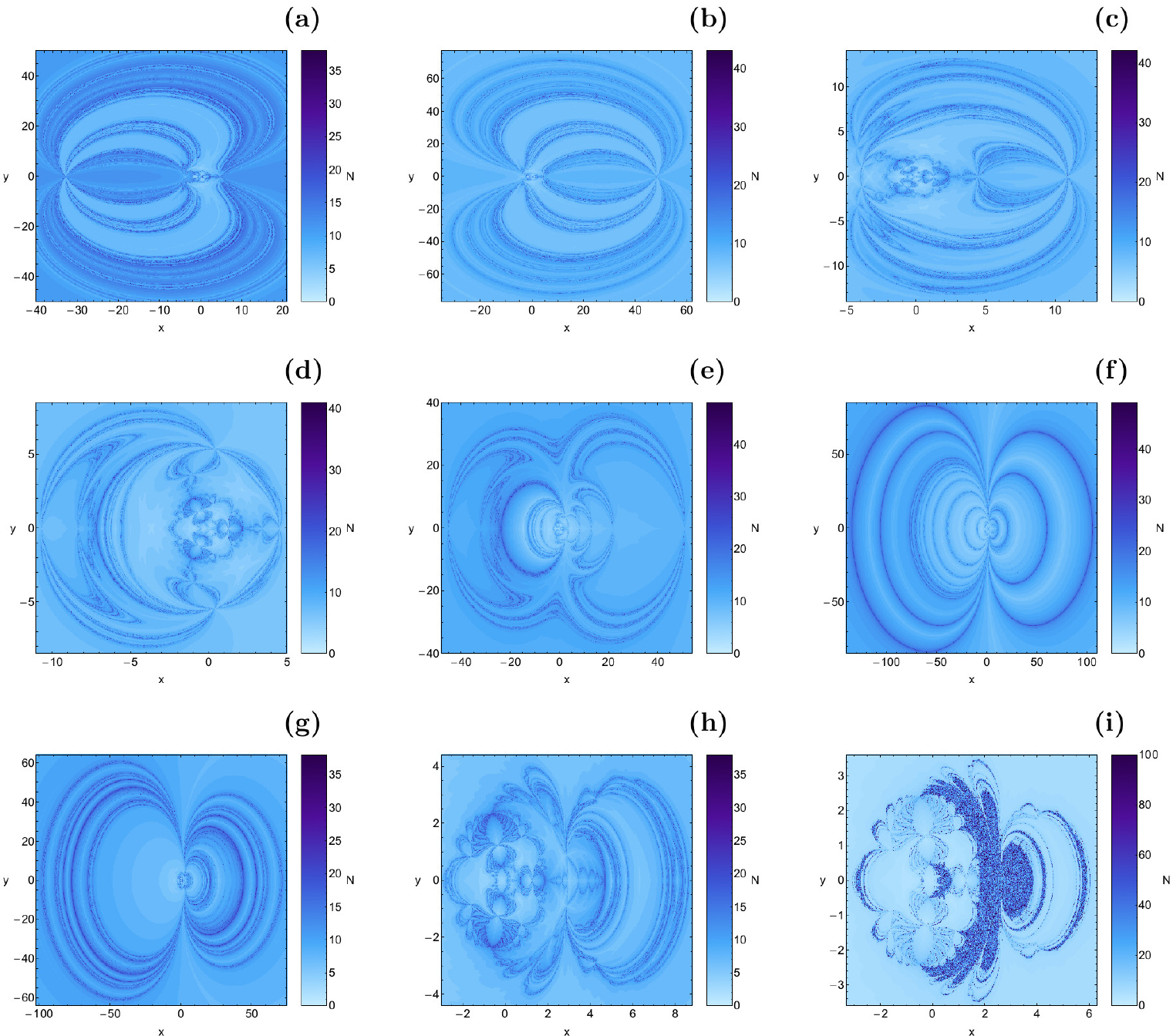}}
\caption{The distribution of the corresponding number $(N)$ of required iterations for obtaining the Newton-Raphson basins of attraction shown in Fig. \ref{pr2}(a-i). The non-converging points are shown in white.}
\label{pr2n}
\end{figure*}

\begin{figure*}[!t]
\centering
\resizebox{\hsize}{!}{\includegraphics{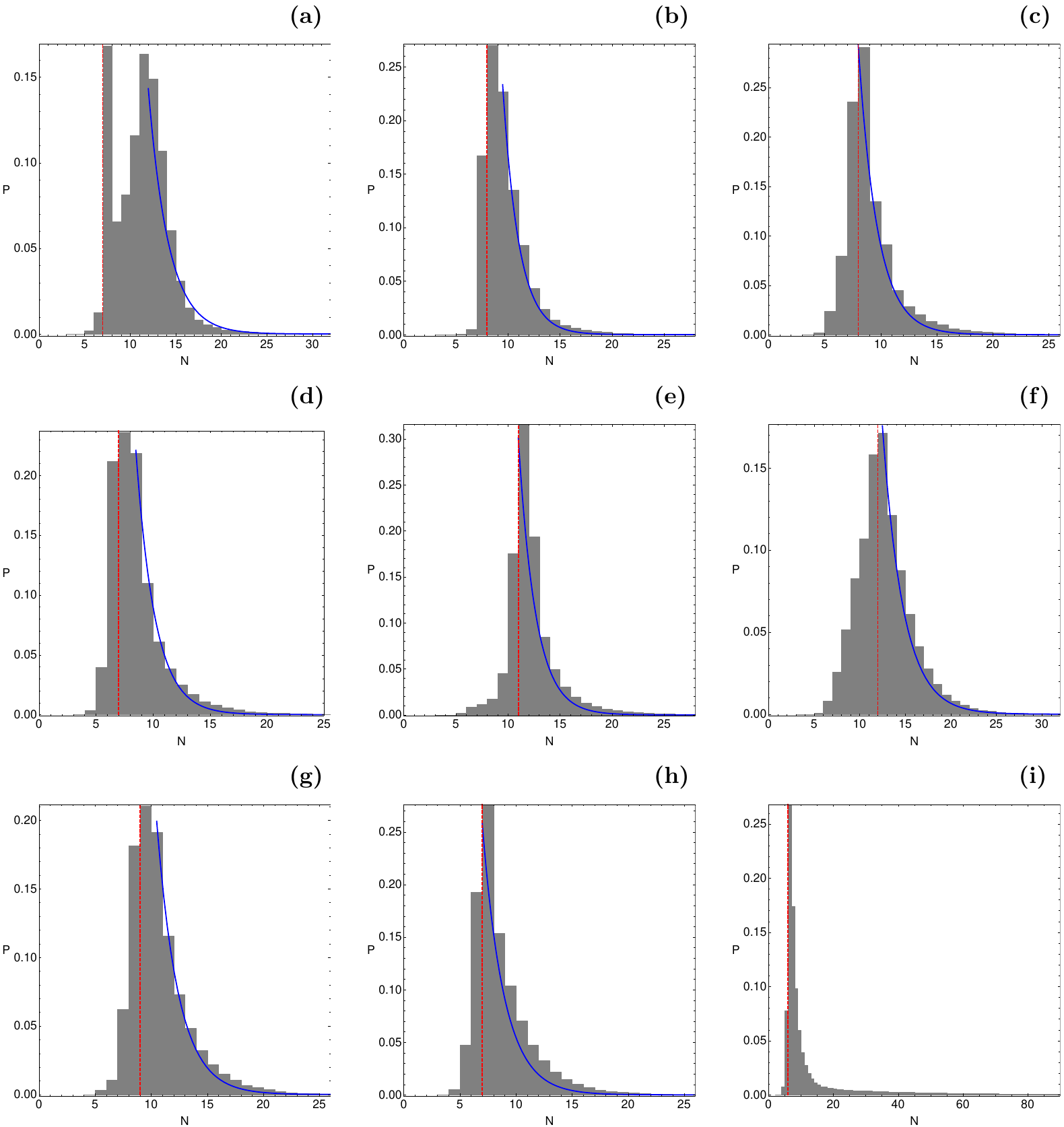}}
\caption{The corresponding probability distribution of required iterations for obtaining the Newton-Raphson basins of attraction shown in Fig. \ref{pr2}(a-i). The vertical dashed red line indicates, in each case, the most probable number $(N^{*})$ of iterations. The blue line is the best fit for the right-hand side $(N > N^{*})$ of the histograms, using a Laplace probability distribution function.}
\label{pr2p}
\end{figure*}

\begin{figure*}[!t]
\centering
\resizebox{\hsize}{!}{\includegraphics{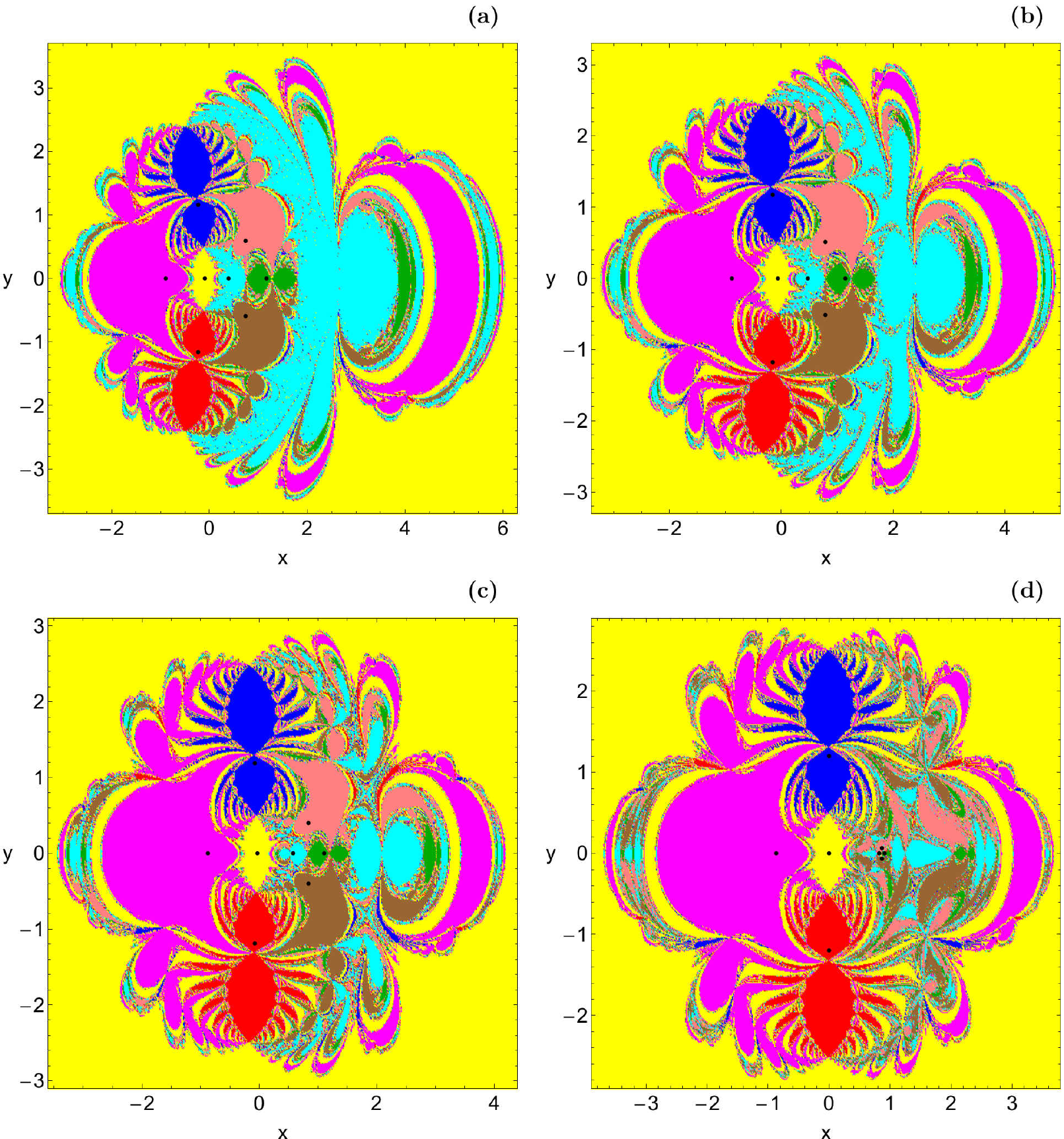}}
\caption{The Newton-Raphson basins of attraction on the configuration $(x,y)$ plane for the third case, where four collinear and four non-collinear equilibrium points are present. (a): $m_3 = 0.4403$; (b): $m_3 = 0.46$; (c): $m_3 = 0.48$; (d): $m_3 = 0.4999$. The positions of the eight equilibrium points are indicated by black dots. The color code, denoting the eight attractors and the non-converging points, is as in Fig. \ref{sm}.}
\label{pr3}
\end{figure*}

\begin{figure*}[!t]
\centering
\resizebox{\hsize}{!}{\includegraphics{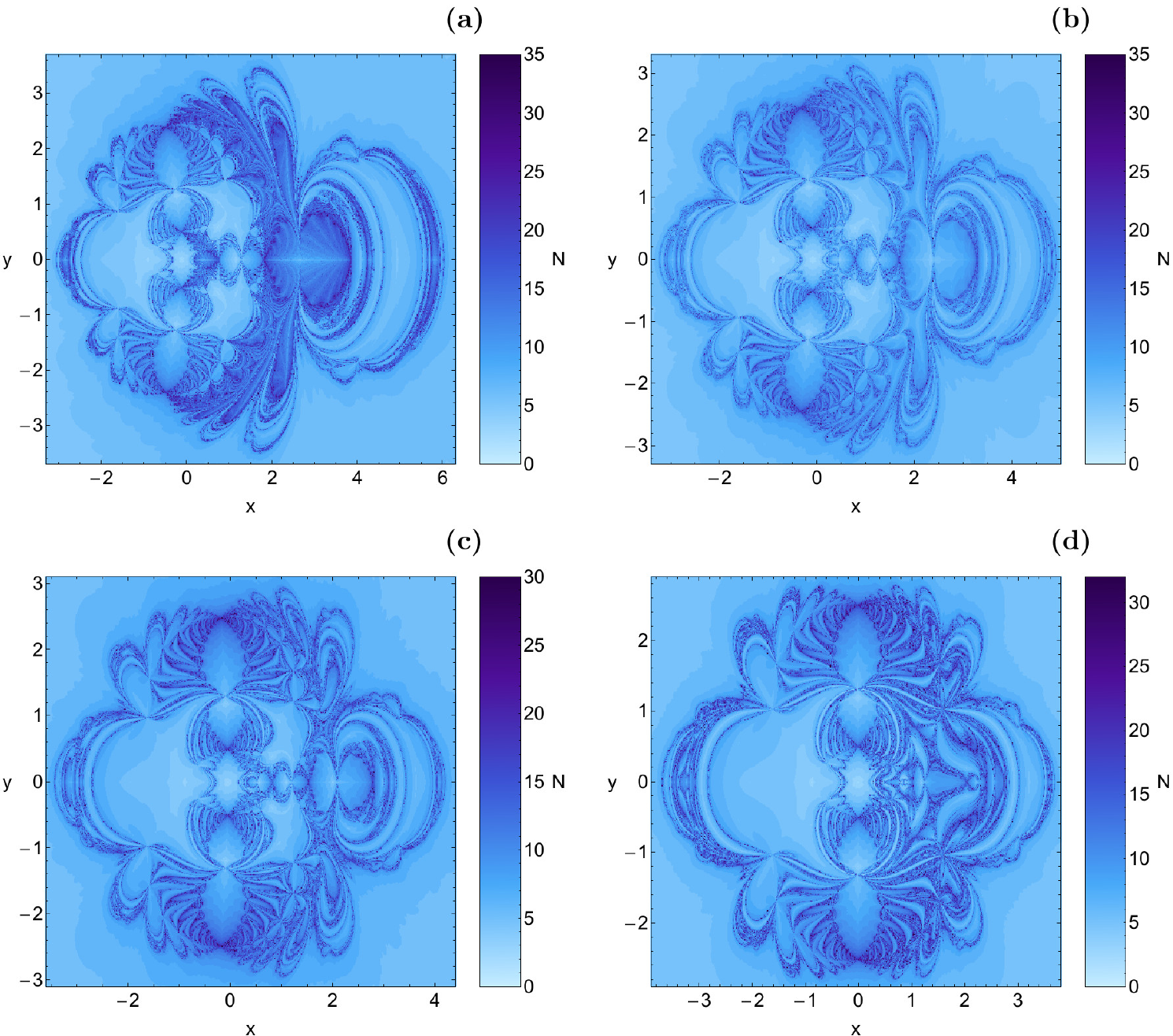}}
\caption{The distribution of the corresponding number $(N)$ of required iterations for obtaining the Newton-Raphson basins of attraction shown in Fig. \ref{pr3}(a-d). The non-converging points are shown in white.}
\label{pr3n}
\end{figure*}

\begin{figure*}[!t]
\centering
\resizebox{\hsize}{!}{\includegraphics{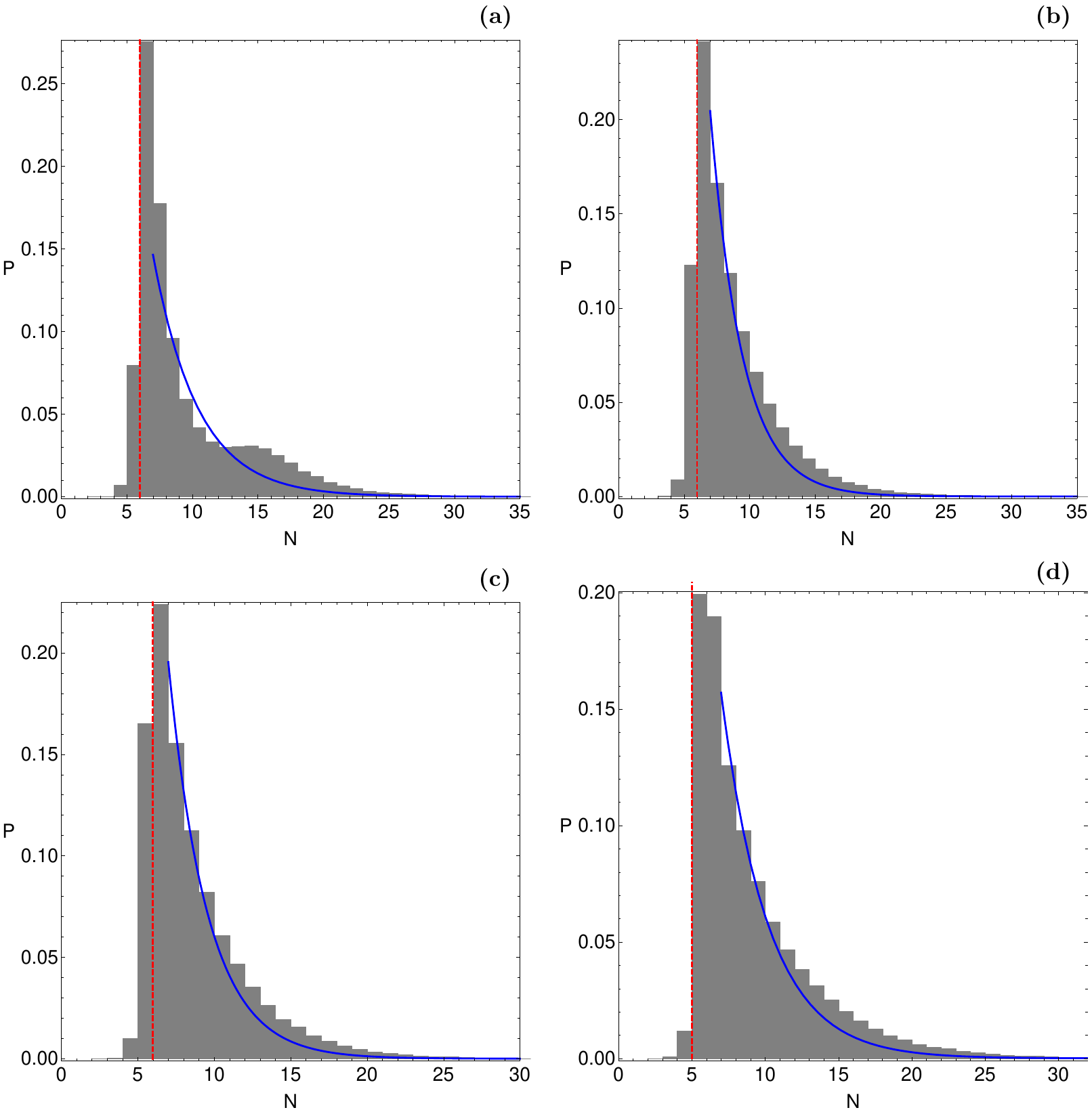}}
\caption{The corresponding probability distribution of required iterations for obtaining the Newton-Raphson basins of attraction shown in Fig. \ref{pr3}(a-d). The vertical dashed red line indicates, in each case, the most probable number $(N^{*})$ of iterations. The blue line is the best fit for the right-hand side $(N > N^{*})$ of the histograms, using a Laplace probability distribution function.}
\label{pr3p}
\end{figure*}

Our investigation begins with the case where two collinear and six non-collinear equilibrium points are present, that is when $0 < m_3 \leq 0.2882761$. In Fig. \ref{pr1} we present a large collection of color-coded plots illustrating the Newton-Raphson basins of convergence for several values of the mass parameter $m_3$. We observe that the existence of one very large primary body and two small ones substantially influences the structure of the attracting basins, with respect to what we seen earlier in Fig. \ref{sm}a where all primaries had equal masses. It is seen that well-formed basins of convergence cover the majority of the configuration $(x,y)$ plane. However, the boundaries of all these basins exhibit a highly fractal\footnote{When we state that a domain displays fractal structure we simply mean that it has a fractal-like geometry however, without conducting, at least for now, any specific calculations for computing the fractal dimensions as in \citet{AVS09}.} structure and we may say that they behave as a ``chaotic sea". The meaning of chaos is justified taking into account that if we choose a starting point $(x_0,y_0)$ inside these fractal areas we will observe that the choice is highly sensitive. In particular, even a slight change in the initial conditions leads to a completely different final destination (different attractor). This implies that in these areas it is almost impossible to predict from which of the libration points each initial condition is attracted by.

As the value of the mass parameter $m_3$ increases the structure of the configuration $(x,y)$ plane changes drastically. We found that the evolution of the structure of the $(x,y)$ plane, with respect to the mass parameter $m_3$, does not follow a specific pattern. On the contrary, we observe sudden and completely unpredicted changes which appear even with a slight change of the value of the mass parameter (see e.g. panels (f) and (g), where $m_3 = 0.258$ and $m_3 = 0.26$, respectively).

Looking at Fig. \ref{pr1}(a-i) one may easily observe a very interesting phenomenon related with the extent of the basins of convergence. Indeed, the extent of the basins of convergence corresponding to non-collinear points is always finite, while on the other hand the extent of the attracting domains of the two collinear points, $L_1$ and $L_4$, extends to infinity. In most of the examined cases it is the $L_1$ basins which is infinite. Additional numerical calculations indicate that for $0 < m_3 \leq 0.2195$, $0.248 \leq m_3 \leq 0.259$, and $0.2745 \leq m_3 \leq 0.276$ the attracting domain with infinite area corresponds to libration point $L_1$, while for $0.2196 \leq m_3 \leq 0.228$ and $0.2725 \leq m_3 \leq 0.274$ the basins of convergence corresponding to collinear point $L_4$ are infinite. For all the other values of $m_3$, always in the range $0 < m_3 < 0.2882761$, it is nearly impossible to know beforehand which of the two collinear equilibrium points dominates with infinite basins of attraction.

In panel (i) of Fig. \ref{pr1}, where $m_3 = 0.2882761$ (or in other words, equal to the first critical value of $m_3$), the basins of convergence are surrounded by a highly chaotic mixture. This mixture is composed of initial conditions of three types: (i) initial conditions attracted by $L_1$; (ii) initial conditions attracted by $L_4$; (iii) initial conditions for which the multivariate Newton-Raphson method does not converge to any of the attractors. These non-converging initial conditions are in reality extremely slow converging points for which the multivariate Newton-Raphson method requires more than 500 iterations. Indeed, if we increase the maximum allowed number of iteration from 500 to 5000 we will see that all non-converging points finally converge, sooner or later, to either $L_1$ or $L_4$.

The distribution of the corresponding number $(N)$ of iterations required for obtaining the desired accuracy is provided in Fig. \ref{pr1n}(a-i), using tones of blue. It is more than evident that initial conditions inside the basins of attraction converge relatively fast $(N < 20)$, while the slowest converging points $(N > 50)$ are those in the vicinity of the basin boundaries. In the same vein, in Fig. \ref{pr1p}(a-i) the corresponding probability distribution of iterations is presented. The definition of the probability $P$ is the following: if we assume that $N_0$ initial conditions $(x_0,y_0)$ converge to one of the available attractors, after $N$ iterations, then $P = N_0/N_t$, where $N_t$ is the total number of initial conditions in every color-coded diagram. The blue lines in the histograms of Fig. \ref{pr1p} indicate the best fit to the right-hand side $N > N^{*}$ of them (more details are given in subsection \ref{geno}).

With increasing value of $m_3$ the the most probable number $(N^{*})$ of iterations is reduced from 33 when $m_3 = 0.0001$ to 11 when $m_3 = 0.22$. For $m_3 > 0.22$ the most probable number of iteration slightly increases up to $m_3 = 0.26$, while for higher values the tendency is reversed and finally for $m_3 = 0.2882761$ we have $N^{*} = 7$. Moreover, it was found that for $m_3 > 0.26$ the range of the probability distribution of the required iterations significantly increases. Indeed, we observe in panel (i) of Fig. \ref{pr1p} that for $m_3 = 0.2882761$ almost the entire range of available iterations, $N \in [0, 500]$, is occupied.

\subsection{Case II: Four collinear points and six non-collinear points}
\label{ss2}

We continue our exploration with the case where four collinear and six non-collinear libration points exist, that is when $0.2882762 \leq m_3 \leq 0.4402$. The Newton-Raphson basins of attraction for nine values of the mass parameter $m_3$ are presented in Fig. \ref{pr2}(a-i). We observe that for values of $m_3$ just above the first critical value $(m_3 = 0.2882761)$ the extent of the basins of attraction corresponding to equilibrium points $L_9$ and $L_{10}$ are much higher than all the other basins except of course for the attracting domain of libration point $L_2$ which extends to infinity. However, as we proceed to higher values of $m_3$ the area of the basins of attraction of equilibrium points $L_9$ and $L_{10}$ slowly decreases and other attractors take over the configuration space.

In this case $(m_3 \in [0.2882762,0.4402])$ all four collinear equilibrium points can have infinite basins of attraction, while those corresponding to non-collinear points always have finite area. Our calculations reveal that each attractor with infinite attracting domain dominate in different range of values of the mass parameter $m_3$. Being more precisely, we found that
\begin{itemize}
  \item Attractor $L_1$ has infinite domains when $0.382 \leq m_3 \leq 0.3945$.
  \item Attractor $L_2$ has infinite domains when $0.2882762 \leq m_3 \leq 0.296$, $0.3806 \leq m_3 \leq 0.3819$, and $0.404 \leq m_3 \leq 0.4402$.
  \item Attractor $L_3$ has infinite domains when $0.2961 \leq m_3 \leq 0.372$, $0.377 \leq m_3 \leq 0.3805$, and $0.396 \leq m_3 \leq 0.403$.
  \item Attractor $L_4$ has infinite domains when $0.3721 \leq m_3 \leq 0.3769$ and $0.3946 \leq m_3 \leq 0.3959$.
\end{itemize}
When the value of $m_3$ approaches the second critical value (0.4402) the area of all finite basins of convergence is reduced and therefore we can easily distinguish, through the color-code diagrams, all the different attracting domains. Indeed, the bug-like structures of the domains corresponding to equilibrium points $L_4$, $L_7$, and $L_8$ and the butterfly-wing shapes of the other basins are again visible. It is interesting to note that in panel (i) of Fig. \ref{pr2}, where $m_3 = 0.4402$, the basins of attraction corresponding to libration points $L_9$ and $L_{10}$ heavily suppress the attracting domains of the collinear point $L_3$.

Our computations suggest that in this range of values of $m_3$ non-converging points are present only when $m_3 = 0.4402$. Even in for this value of the mass parameter however, the relative fraction of non-converging points is so small $(< 0.01\%)$ that their presence is almost negligible. Once more, these non-converging initial conditions are in reality extremely slow converging points with $N > 500$.

In Fig. \ref{pr2n}(a-i) we illustrate the distribution of the corresponding number $(N)$ of iterations required for obtaining the desired accuracy. Looking at panel (i) of Fig. \ref{pr2n} we may observe a very strange phenomenon. So far, the lowest required number of iterations have been identified for initial conditions inside the basins of attraction. However for $m_3 = 0.4402$ (see panel (i) of Fig. \ref{pr2n}) it is seen all the initial conditions which are attracted either by $L_9$ or $L_{10}$ possess the highest numbers of iteration $(N > 50)$ observed in this case. At the moment, we cannot give a logical explanation for this phenomenon (the required number of iterations of initial conditions inside the basins of convergence to be higher than the iterations of the initial conditions in the vicinity of the basin boundaries) which still remains a mystery. Perhaps the fact that this case corresponds to the second critical value of the mass parameter $m_3 = 0.4402$ gives an explanation to this strange and bizarre behavior.

The corresponding probability distribution of iterations is given in Fig. \ref{pr2p}(a-i). The the most probable number $(N^{*})$ of iteration starts at 7 for $m_3 = 0.2882762$, then it increases up to $m_3 = 0.38$, while for $m_3 > 0.38$ it decreases until it reaches the lowest observed value, $N^{*} = 6$, for $m_3 = 0.4402$.

\subsection{Case III: Four collinear points and four non-collinear points}
\label{ss3}

The last case under consideration concerns the scenario according to which the PERFBP with two equal masses has four collinear and four non-collinear equilibrium points $(0.4403 \leq m_3 < 1/2)$. In Fig. \ref{pr3}(a-d) we provide the Newton-Raphson basins of convergence for four values of the mass parameter $m_3$. In this range of values of $m_3$ the changes on the configuration $(x,y)$ plane, due to the variation of the mass parameter, are not so prominent as in the previous two cases. Thus we decided to present in Fig. \ref{pr3} only four (instead of nine) color-coded convergence diagrams.

\begin{figure*}[!t]
\centering
\resizebox{\hsize}{!}{\includegraphics{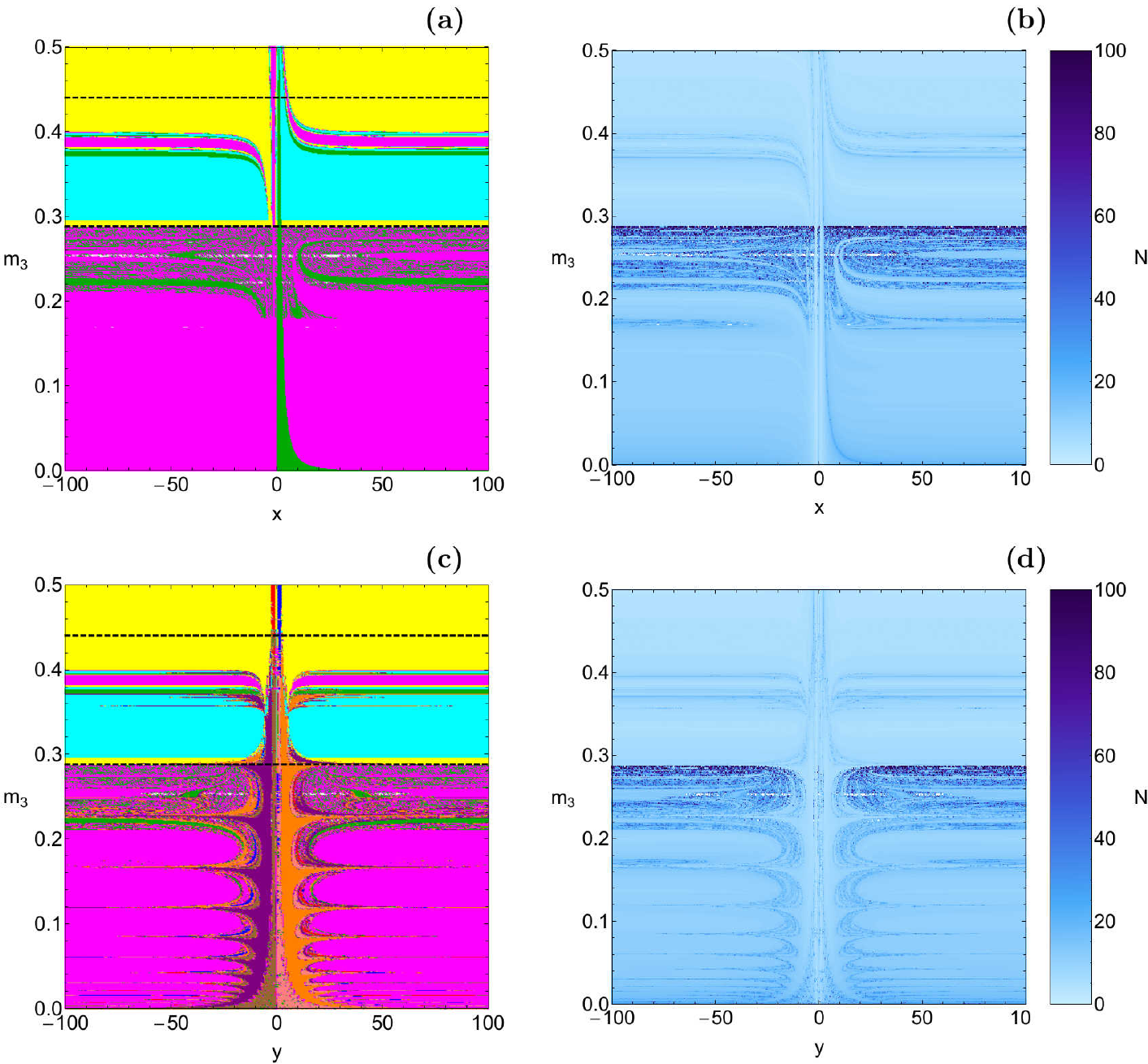}}
\caption{The Newton-Raphson basins of attraction on the (a-upper left): $(x,m_3)$ plane and (c-lower left): $(y,m_3)$ plane, when $m_3 \in (0,1/2)$. The color code denoting the attractors is the same as in Fig. \ref{sm}. The horizontal black dashed lines indicate the two critical values of $m_3$. Panels (b) and (d): The distribution of the corresponding number $(N)$ of required iterations for obtaining the Newton-Raphson basins of attraction shown in panels (a) and (c), respectively.}
\label{xym3}
\end{figure*}

\begin{figure*}[!t]
\centering
\resizebox{\hsize}{!}{\includegraphics{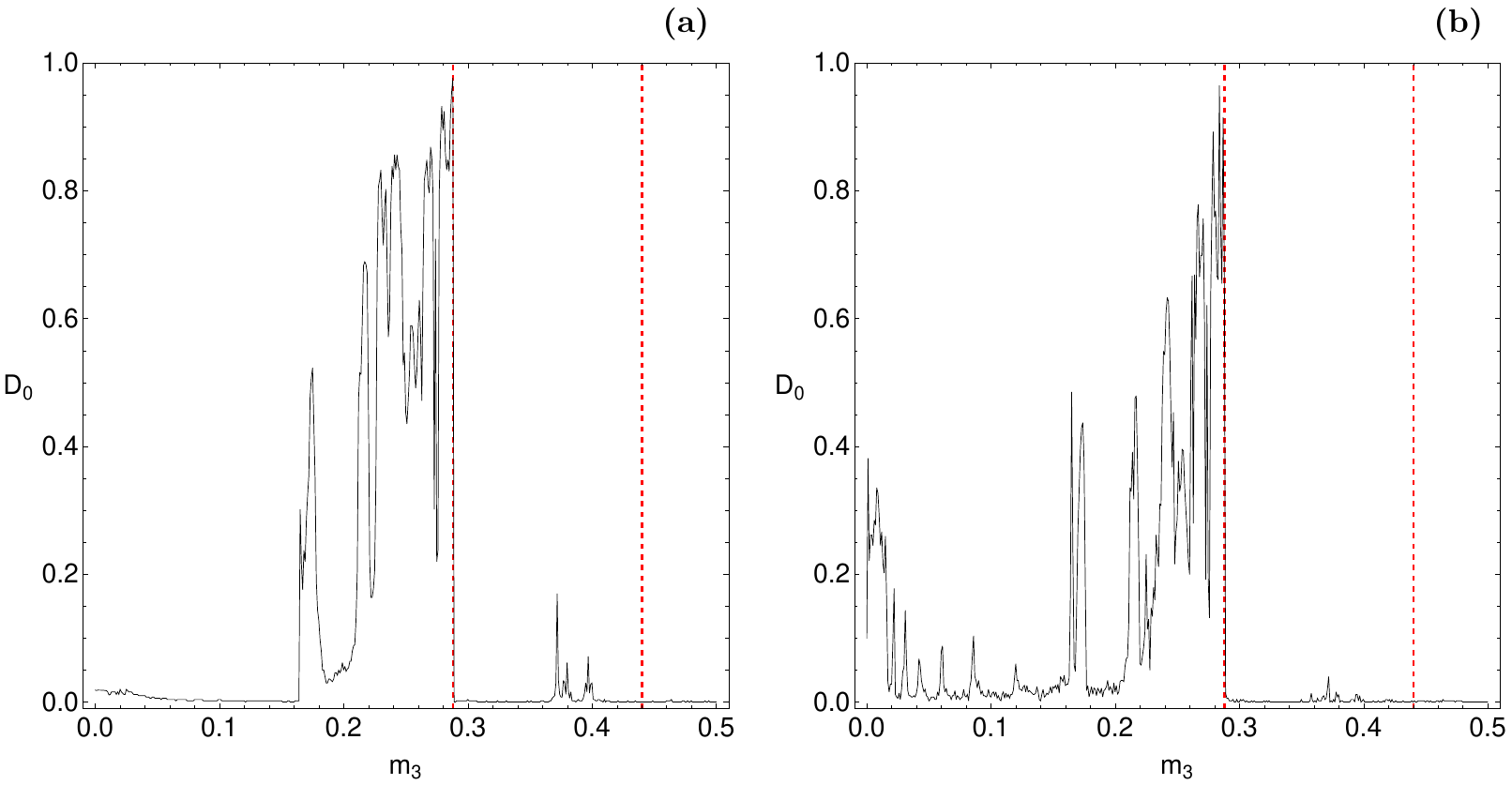}}
\caption{Evolution of the fractal dimension $D_0$ of the (a-left): $(x,m_3)$ plane and (b-right): $(y,m_3)$ plane of Figs. \ref{xym3} as a function of the mass parameter $m_3$. $D_0 = 1$ means total fractality, while $D_0 = 0$ implies zero fractality. The red dashed vertical lines indicate the two critical values of the mass parameter $m_3$, which distinguish between the three cases regarding the total number and the type of the equilibrium points.}
\label{frac}
\end{figure*}

As it is seen in Fig. \ref{lgs} for $m_3 > 0.4402$ the pair of the non-collinear libration points $L_9$ and $L_{10}$ disappears. If we compare the color-coded diagrams shown in panel (i) of Fig. \ref{pr2} and in panel (a) of Fig. \ref{pr3} we will see that their structure is almost identical with only one major difference: for $m_3 = 0.4403$ the $L_3$ attractor has completely assimilated the basins of convergence of equilibrium points $L_9$ and $L_{10}$, which are present for $m_3 = 0.4402$. As the value of the mass parameter tends asymptotically to 1/2 the following phenomena take place in the configuration $(x,y)$ plane
\begin{itemize}
  \item The area of the attracting domains corresponding to libration points $L_1$, $L_7$, and $L_8$ increases.
  \item The area of the basins of converging of the attractors $L_5$ and $L_6$ seems almost unperturbed.
  \item The area of the basins of attractions corresponding to $L_3$ and $L_4$ decreases.
  \item The convergence domain of collinear point $L_2$ is infinite. In fact, in this case $(0.4403 \leq m_3 < 1/2)$ only the attractor $L_2$ has infinite attracting domain.
  \item The shape (bug-like or butterfly wing-like) of all the basins of attraction remains almost unperturbed.
  \item In this case, there is no numerical evidence of non-converging initial conditions, whatsoever.
\end{itemize}

Looking at panel (d) of Fig. \ref{pr3}, where $m_3 = 0.4999$, we can numerically verify a statement we made earlier in Section 3. We see that the equilibrium points $L_3$, $L_4$, $L_5$, and $L_6$ are very close to one another, while they tend to collide to the limit $m_3 \to 1/2$. Now if we plot the corresponding four basins of convergence with the same color we will obtain a color-code diagram almost identical to that presented in panel (f) of Fig. 3 in \citet{Z16a}, while the only difference will be its orientation (a rotation by 90$^{\circ}$).

The corresponding number $(N)$ of required iterations for the desired accuracy is shown in Fig. \ref{pr3n}(a-d), while the probability distribution of iterations is presented in Fig. \ref{pr3p}(a-d). It was found, that for more than 95\% of the examined initial conditions on the configuration $(x,y)$ plane the iterative scheme (\ref{nrm}) needs no more than 30 iterations for obtaining the desired accuracy. Furthermore, the average value of required number $(N)$ of iterations remains almost constant, $N^{*} = 6$, throughout the range of values of $m_3$. Only for $m_3 = 0.4999$ it was measured a little bit lower at $N^{*} = 5$.

\subsection{An overview analysis}
\label{geno}

The color-coded convergence diagrams on the configuration $(x,y)$ space, presented earlier in Figs. \ref{pr1}, \ref{pr2}, and \ref{pr3}, provide sufficient information regarding the attracting domains however for only a fixed value of the mass parameter $m_3$. In order to overcome this handicap we can define another type of initial conditions which will allow us to scan a continuous spectrum of $m_3$ values rather than few discrete levels. The most easy configuration is to set on the two coordinates $(x,y)$ equal to zero, while the initial value of the other coordinate will vary in the interval $[-100,100]$. This technique allows us to construct, once more, a two-dimensional plane in which the $x$ or the $y$ coordinate is the abscissa, while the value of $m_3$ is always the ordinate. In panel (a) of Fig. \ref{xym3} we present the attracting domains of the $(x,m_3)$ plane when $m_3 \in (0,1/2)$, while in panel (b) of the same figure the distribution of the corresponding number $(N)$ of required iterations for obtaining the Newton-Raphson basins of attraction is shown. In panels (c) and (d) of Fig. \ref{xym3} the corresponding results of the $(y,m_3)$ plane are shown. In both cases, the two critical values of $m_3$, which delimit the three cases, are indicated using horizontal black dashed lines.

It is interesting to note that when $0.235 < m_3 < 0.2882$ a highly chaotic layer is present in both types of diagrams. Therefore, in this interval the task of knowing beforehand which of the collinear equilibrium points has infinite basins of attraction is next to impossible. Looking at panels (b) and (d) of Fig. \ref{xym3} it becomes more than evident that the vast majority of initial conditions in this range of values of $m_3$ do converge to one of the attractors however, after a relatively high number of iterations $(N > 50)$. Finally, it should be noted that for about $m_3 = 0.258$ we detected a small portion of non-converging points. Additional numerical calculations indicated that these non-converging points are in reality extremely slow converging points which need more than 500 of iterations in order to reach to one the attractors. A similar behaviour was also observed for the non-converging points in subsections \ref{ss1} and \ref{ss2}. In general terms we may say that in all other regions apart from the interval $m_3 \in (0.235, 28882)$ the multivariate Newton-Raphson method requires no more than 30 iterations in order to converge to one of the attractors.

So far we have discussed the fractality of the several two-dimensional planes only in a qualitative way. More precisely, we seen that the highly fractal areas are those in which we cannot predict from which attractor (equilibrium point) each initial condition is attracted. On the other hand, inside the basins of convergence the degree of fractality is zero and the final state of the initial conditions is well known and of course predictable. At this point we shall provide a quantitative analysis regarding the degree of fractality for the $(x,m_3)$ and $(y,m_3)$ planes shown earlier in panels (a) and (c) of Fig. \ref{xym3}, respectively. In order to measure the degree of fractality we have computed the uncertainty dimension \citep{O93} for different values of the mass parameter $m_3$, thus following the computational method introduced in \citet{AVS01}. Obviously, this degree of fractality is completely independent of the initial conditions we used to compute it.

The evolution of the uncertainty dimension $D_0$ for both $(x,m_3)$ and $(y,m_3)$ planes, as a function of the mass parameter $m_3$, is shown in Fig. \ref{frac}(a-b). The computation of the uncertainty dimension was done for only a ``1D slice'' of initial conditions of Figs. \ref{xym3}, and for that reason $D_0 \in (0,1)$. It is interesting to note that in both types of planes for the first critical value of $m_3$, that is $m_3 = 0.2882761$, the uncertainty dimension tends to one. This means that for that critical value there is a total fractalization of the corresponding planes and the chaotic set becomes ``dense" in the limit. On the contrary, for the second critical value $(m_3 = 0.4402)$ the uncertainty dimension is almost zero. Looking carefully both panels of Fig. \ref{xym3} we may conclude that the highest degree of fractality is observed when $0.16 < m_3 < 0.18$ and especially when $0.2 < m_3 < 0.2882761$. On the other hand the lowest degree of fractality is measured in the third case $(m_3 > 0.4402)$, where four collinear and four non-collinear equilibrium points exist.

Before closing this numerical investigation we would like to shed some light to the probability distributions of iterations presented in Figs. \ref{pr1p}, \ref{pr2p}, and \ref{pr3p}. In particular, it would be very interesting to try to obtain the best fit of the tails\footnote{By the term ``tails" of the distributions we refer to the right-hand side of the histograms, that is, for $N > N^{*}$.} of the distributions. For finding the best fit of the tails we tried several single types of distributions (Laplace, Maxwell-Boltzmann, Rayleigh, Pascal, Poisson, etc). Our calculations strongly indicate that in the vast majority of the cases the Laplace distribution is the best fit to our data. The only two cases where the Laplace distribution fails to fit the corresponding numerical data are the cases corresponding to the two critical values of the mass parameter $m_3$ (see panels (i) in Figs. \ref{pr1p} and \ref{pr2p}).

The probability density function (PDF) of the Laplace distribution is given by
\begin{equation}
P(N | \mu,b) = \frac{1}{2b}
 \begin{cases}
      \exp\left(- \frac{\mu - N}{b} \right), & \text{if } N < \mu \\
      \exp\left(- \frac{N - \mu}{b} \right), & \text{if } N \geq \mu
 \end{cases},
\label{pdf}
\end{equation}
where $\mu$ is the location parameter, while $b > 0$, is the diversity. In our case we are interested only for the $x \geq \mu$ part of the distribution function.

In Table \ref{table1} we present the values of the location parameter $\mu$ and the diversity $b$, as they have obtained through the best fit, for all cases discussed in Figs. \ref{pr1p}, \ref{pr2p}, and \ref{pr3p}. One may observe that for most of the cases the location parameter $\mu$ is very close to the most probable number $N^{*}$ of iterations, while in some cases these two quantities coincide. Here we would like to emphasize that the Laplace distribution is only a first good approximation to our data. Additional numerical calculations indicate that if we use a mixture of several types of distributions, instead of a single type of distribution (i.e., the Laplace distribution), the fit is much better. However we feel that this task is out of the scope and the spirit of this paper and therefore we did not pursue it.

\begin{table}[!ht]
\begin{center}
   \caption{The values of the location parameter $\mu$ and the diversity $b$, related to the most probable number $N^{*}$ of iterations, for all the studied cases shown in Figs. \ref{pr1p}, \ref{pr2p}, and \ref{pr3p}.}
   \label{table1}
   \setlength{\tabcolsep}{10pt}
   \begin{tabular}{@{}lrrrr}
      \hline
      Figure & $m_3$ & $N^{*}$ & $\mu$ & $b$ \\
      \hline
      \ref{pr1p}a &    0.0001 & 33 & $N^{*}$     &  5.09905946 \\
      \ref{pr1p}b &      0.01 & 17 & $N^{*} + 1$ &  3.22370099 \\
      \ref{pr1p}c &      0.05 & 13 & $N^{*}$     &  2.86554309 \\
      \ref{pr1p}d &      0.22 & 10 & $N^{*} + 2$ &  3.58862734 \\
      \ref{pr1p}e &      0.25 & 12 & $N^{*} + 2$ &  7.46437951 \\
      \ref{pr1p}f &     0.258 & 13 & $N^{*} + 2$ & 12.84964306 \\
      \ref{pr1p}g &      0.26 & 16 & $N^{*} + 5$ & 11.56009567 \\
      \ref{pr1p}h &     0.275 & 10 & $N^{*} + 6$ & 17.09014571 \\
      \ref{pr1p}i & 0.2882761 &  7 &           - &           - \\
      \hline
      \ref{pr2p}a & 0.2882762 &  7 & $N^{*} + 4$ & 2.22864018 \\
      \ref{pr2p}b &      0.29 &  8 & $N^{*} + 1$ & 1.55241568 \\
      \ref{pr2p}c &      0.30 &  8 & $N^{*}$     & 1.70743943 \\
      \ref{pr2p}d &      0.35 &  7 & $N^{*} + 1$ & 1.68094743 \\
      \ref{pr2p}e &      0.37 & 11 & $N^{*}$     & 1.66194465 \\
      \ref{pr2p}f &      0.38 & 12 & $N^{*}$     & 2.28836276 \\
      \ref{pr2p}g &      0.40 &  9 & $N^{*} + 1$ & 1.94075381 \\
      \ref{pr2p}h &      0.42 &  7 & $N^{*}$     & 1.93663675 \\
      \ref{pr2p}i &    0.4402 &  6 &           - &          - \\
      \hline
      \ref{pr3p}a & 0.4403 &  6 & $N^{*} + 1$ & 3.41485683 \\
      \ref{pr3p}b &   0.46 &  6 & $N^{*} + 1$ & 2.44498156 \\
      \ref{pr3p}c &   0.48 &  6 & $N^{*} + 1$ & 2.55876653 \\
      \ref{pr3p}d & 0.4999 &  5 & $N^{*} + 2$ & 3.18349760 \\
      \hline
   \end{tabular}
\end{center}
\end{table}

\section{Concluding remarks}
\label{conc}

The scope of this research paper was to numerically determine the basins of convergence associated with the equilibrium points. In the PERFBP with two equal masses the number, the position and the type of the libration points strongly depends on the value of the mass parameter $m_3$. With the help of the multivariate version of the Newton-Raphson iterative scheme we managed to unveil the extraordinary and magnificent structures of the basins of attraction corresponding to the equilibrium points of the dynamical system. These basins play an important role as they describe how each point on the configuration $(x,y)$ plane is attracted by the libration points which act as attractors. Our numerical exploration revealed how the position of the equilibrium points and of course the structure of the attracting areas are influenced by the mass parameter $m_3$. Furthermore, we related the several basins of attraction with the corresponding distribution of the required number of iterations. To our knowledge, this is the first time that such a thorough and systematic numerical investigation, regarding the basins of attraction, takes place in the PERFBP and this is exactly the novelty as well as the importance of the current work.

The main results of our numerical research are the following:
\begin{enumerate}
  \item We observed that the change on the value of the mass parameter $m_3$ mostly influences the shape and the geometry of the basins of attraction when $0 < m_3 < 0.4402$. Indeed, in this range it is almost impossible to know beforehand the structure of the attracting regions, as even a tiny change on the value of $m_3$ leads to a complete different structure on the $(x,y)$ plane. For larger values of $m_3$ the influence of the same parameter on the attracting regions is much more milder.
  \item In all examined cases the area of the basins of convergence corresponding to the non-collinear libration points is finite. On the other hand, all four collinear equilibrium points can have infinite attracting domains. In particular, it was found that when $0.2882762 \leq m_3 \leq 0.4402$ each collinear attractor dominate in different ranges of the values of $m_3$.
  \item The iterative method was found to converge very fast $(0 \leq N < 15)$ for initial conditions around each equilibrium point, fast $(15 \leq N < 25)$ and slow $(25 \leq N < 50)$ for initial conditions that complement the central regions of the very fast convergence, and very slow $(N \geq 50)$ for initial conditions of dispersed points lying either in the vicinity of the basin boundaries, or between the dense regions of the equilibrium points.
  \item In general terms we concluded that the average value of required iterations $(N^{*})$ for obtaining the desired accuracy decreases with increasing value of the mass parameter. In almost all cases, the Newton-Raphson method, for more than 90\% of the initial conditions, requires less than 60 iterations to converge to one of the available attractors.
  \item Our calculations strongly suggest that non-converging points on the configuration $(x,y)$ plane exist only when $0 < m_3 < 0.4402$. Being more precisely, they appear mostly for values of $m_3$ just before the two critical values ($m_3 = 0.2882761$ and $m_3 = 0.4402$). A deeper analysis on these points revealed the fact that these points are not true non-converging points. In reality they are extremely slow converging points which require a huge number of iterations $(N > 500)$ in order to reach to one of the attractors.
  \item In the case where eight equilibrium points exist (two collinear and six non-collinear) we observed the highest degree of fractality, especially when $0.2 < m_3 < 0.2882761$. On the other hand, in the third case $(m_3 > 0.4402)$, where four collinear and four non-collinear equilibrium points exist, we measured the lowest degree of fractality.
  \item Our tests indicate that our numerical data, corresponding to the histograms with the probability distributions of the required iterations, are best fitted by the Laplace probability distribution function (PDF). Only the cases just before the two critical values of the mass parameter (which have long tails) cannot be fitted well by a Laplace PDF.
\end{enumerate}

A double precision code, written in standard \verb!FORTRAN 77!, has been deployed for performing all the required numerical calculations regarding the basins of convergence. For the graphical illustration of the paper, we used the latest version 11.0 of Mathematica$^{\circledR}$ \cite{W03}. For the classification of each set of the initial conditions on the several types of two-dimensional planes, we needed about 5 minutes of CPU time using a Quad-Core i7 2.4 GHz PC, depending of course on the required number of iterations. When an initial condition had converged to one of the attractors with the predefined accuracy the iterative procedure was effectively ended and proceeded to the next available initial condition.

Judging by the novel results revealed through the detailed and systematic numerical exploration we believe that we successfully completed our computational task. We hope that our investigation and the corresponding outcomes to be useful in the field of attracting domains in the PERFBP. Taking into account that the current analysis of the case with two equal masses was encouraging it is our future plans to study the general case with three unequal masses. In addition, it would be highly interesting to try and use other types of iterative formulae (of higher order with respect to the classical Newton-Raphson method) and compare the similarities as well as the differences of the structures of the basins of attraction.

\section*{Acknowledgments}

I would like to express my warmest thanks to the anonymous referee for the careful reading of the manuscript and for all the apt suggestions and comments which allowed us to improve both the quality and the clarity of the paper.


\begin{thebibliography}{}
\footnotesize

\bibitem[\protect\citeauthoryear{Aguirre et al.}{2001}]{AVS01} Aguirre, J., Vallejo, J.C., Sanju\'{a}n, M.A.F.: Wada basins and chaotic invariant sets in the H\'{e}non-Heiles system. Phys. Rev E \textbf{64}, 066208-1--11 (2001)

\bibitem[\protect\citeauthoryear{Aguirre et al.}{2009}]{AVS09} Aguirre, J., Viana, R.L., Sanju\'{a}n, M.A.F.: Fractal structures in nonlinear dynamics, Rev. Mod. Phys. \textbf{81}, 333-386 (2009)

\bibitem[\protect\citeauthoryear{\'{A}lvarez-Ram\'{i}rez et al.}{2015a}]{AB15} \'{A}lvarez-Ram\'{i}rez M., Barrab\'{e}s E.: Transport orbits in an equilateral restricted four-body problem, Celest. Mech. Dyn. Astron. \textbf{121}, 191-210 (2015)

\bibitem[\protect\citeauthoryear{\'{A}lvarez-Ram\'{i}rez et al.}{2015b}]{ASS15} \'{A}lvarez-Ram\'{i}rez M., Skea J.E.F., Stuchi T.J.: Nonlinear stability analysis in a equilateral restricted four-body problem. Astrophys. Space Sci. \textbf{358}, 3 (2015)

\bibitem[\protect\citeauthoryear{Baltagiannis \& Papadakis}{2011a}]{BP11a} Baltagiannis, A.N., Papadakis, K.E.: Equilibrium points and their stability in the restricted four-body problem. Int. J. Bifurc. Chaos \textbf{21}, 2179-2193 (2011a)

\bibitem[\protect\citeauthoryear{Baltagiannis \& Papadakis}{2011b}]{BP11b} Baltagiannis, A.N., Papadakis, K.E.: Families of periodic orbits in the restricted four-body problem. Astrophys. Space Sci. \textbf{336}, 357-367 (2011b)

\bibitem[\protect\citeauthoryear{Baltagiannis \& Papadakis}{2013}]{BP13} Baltagiannis, A.N., Papadakis, K.E.: Periodic solutions in the Sun-Jupiter-Trojan Asteroid-Spacecraft system. Planet. Space Sci. \textbf{75}, 148-157 (2013)

\bibitem[\protect\citeauthoryear{Burgos-Garc\'{i}a \& Delgado}{2013}]{BGD13} Burgos-Garc\'{i}a, J., Delgado, J.: Periodic orbits in the restricted four-body problem with two equal masses, Astrophys. Space Sci. \textbf{345}, 247-263 (2013)

\bibitem[\protect\citeauthoryear{Ceccaroni \& Biggs}{2010}]{CB10} Ceccaroni, M., Biggs, J.: Extension of low-thrust propulsion to the autonomous coplanar circular restricted four body problem with application to future Trojan Asteroid missions. In: 61st International Astronautical Congress, IAC 2010, Prague, Czech Republic (2010)

\bibitem[\protect\citeauthoryear{Croustalloudi \& Kalvouridis}{2007}]{CK07} Croustalloudi, M., Kalvouridis, T.: Attracting domains in ring-type N-body formations. Planet. Space Sci. \textbf{55}, 53-69 (2007)

\bibitem[\protect\citeauthoryear{Croustalloudi \& Kalvouridis}{2013}]{CK13} Croustalloudi, M.N., Kalvouridis, T.J.: The Restricted 2+2 body problem: Parametric variation of the equilibrium states of the minor bodies and their attracting regions, ISRN Astronomy and Astrophysics, Article ID 281849 (2013)

\bibitem[\protect\citeauthoryear{de Almeida Prado}{2005}]{dAP05} de Almeida Prado, A.F.B.: Numerical and analytical study of the gravitational capture in the bicircular problem. Adv. Sp. Res. \textbf{36} 578-584 (2005)

\bibitem[\protect\citeauthoryear{Douskos}{2010}]{D10} Douskos, C.N.: Collinear equilibrium points of Hill's problem with radiation and oblateness and their fractal basins of attraction. Astrophys. Space Sci. \textbf{326}, 263-271 (2010)

\bibitem[\protect\citeauthoryear{Gousidou-Koutita \& Kalvouridis}{2009}]{GKK09} Gousidou-Koutita, M., Kalvouridis, T.J.: On the efficiency of Newton and Broyden numerical methods in the investigation of the regular polygon problem of $(N + 1)$ bodies. Appl. Math. Comput. \textbf{212}, 100-112 (2009)

\bibitem[\protect\citeauthoryear{Hadjidemetriou}{1980}]{H80} Hadjidemetriou, J.D.: The restricted planetary 4-body problem. Celest. Mech. \textbf{21}, 63-71 (1980)

\bibitem[\protect\citeauthoryear{Jorba}{2000}]{J00} Jorba, A.: On practical stability regions for the motion of a small particle close to the equilateral points of the real Earth-Moon system, in: Proceedings of the 3rd International Symposium (HAMSYS-98) Heldat P\'{a}tzcuaro, World Scientiﬁc Monograph Series in Mathematics, vol. 6, World Scientiﬁc, River Edge, NJ, USA, pp. 197-213 (2000)

\bibitem[\protect\citeauthoryear{Kalvouridis}{2008}]{K08} Kalvouridis, T.J.: On some new aspects of the photo-gravitational Copenhagen problem. Astrophys. Space Sci. \textbf{317}, 107-117 (2008)

\bibitem[\protect\citeauthoryear{Kalvouridis \& Gousidou-Koutita}{2012}]{KGK12} Kalvouridis, T.J., Gousidou-Koutita, M.Ch.: Basins of attraction in the Copenhagen problem where the primaries are magnetic dipoles. Applied Mathematics \textbf{3}, 541-548 (2012)

\bibitem[\protect\citeauthoryear{Kloppenborg}{2010}]{KSM10} Kloppenborg, B., Stencel, R., Monnier, J.D., Schaefer, G., Zhao, M., Baron, F., McAlister, H., et al.: Infrared images of the transiting disk in the epsilon Aurigae system. Nat. Lett. \textbf{464}, 870-872 (2010)

\bibitem[\protect\citeauthoryear{Kumari \& Kushvah}{2014}]{KK14} Kumari, R., Kushvah, B.S.: Stability regions of equilibrium points in restricted four-body problem with oblateness effects. Astrophys. Space Sci. \textbf{349}, 693-704 (2014)

\bibitem[\protect\citeauthoryear{Leandro}{2006}]{L06} Leandro, E.S.G.: On the central congurations of the planar restricted four-body problem, J. Differ. Equ. \textbf{226}, 323-351 (2006)

\bibitem[\protect\citeauthoryear{Machuy et al.}{2007}]{MdS07} Machuy, A.L., de Almeida Prado, A.F.B., Stuchi, T.J.: Numerical study of the time required for the gravitational capture in the bi-circular four-body problem. Adv. Sp. Res. \textbf{40} 118-124 (2007)

\bibitem[\protect\citeauthoryear{Marchal}{1990}]{M90} Marchal, C.: The Three-Body Problem. Elsevier, New York (1990)

\bibitem[\protect\citeauthoryear{Melita et al.}{2008}]{MLJ08} Melita, M.D., Licandro, J., Jones, D.C., Williams, I.P.: Physical properties and orbital stability of the Trojan asteroids. Icarus \textbf{195}, 686-697 (2008)

\bibitem[\protect\citeauthoryear{Michalodimitrakis}{1981}]{M81} Michalodimitrakis, M.: The circular restricted four-body problem, Astrophys. Space Sci. \textbf{75}, 289-305 (1981)

\bibitem[\protect\citeauthoryear{Moulton}{1900}]{M00} Moulton, F.R.: On a class of particular solutions of the problem of four bodies. Trans. Amer. Math. Soc. \textbf{1}, 17-29 (1900)

\bibitem[\protect\citeauthoryear{Ott}{1993}]{O93} Ott, E: Chaos in Dynamical Systems. Cambridge University Press, Cambridge (1993)

\bibitem[\protect\citeauthoryear{Papadakis}{2007}]{P07} Papadakis, K.E.: Asymptotic orbits in the restricted four-body problem, Planet. Space Sci. \textbf{55}, 1368-1379 (2007)

\bibitem[\protect\citeauthoryear{Papadouris \& Papadakis}{2013}]{PP13} Papadouris, J.P., Papadakis, K.E.: Equilibrium points in the restricted four-body problem, Astrophys. Space Sci. \textbf{344}, (21-38 (2013)

\bibitem[\protect\citeauthoryear{Robutel \& Gabern}{2006}]{RG06} Robutel, P., Gabern, F.: The resonant structure of Jupiter's Trojan asteroids-I. Long-term stability and diffusion. Mon. Not. R. Astron. Soc. \textbf{372}, 1463-1482 (2006)

\bibitem[\protect\citeauthoryear{Scharz et al.}{2009a}]{SSD09a} Schwarz, R., S\"{u}li, \`{A}., Dvorac, R., Pilat-Lohinger, E.: Stability of Trojan planets in multi-planetary systems. Celest. Mech. Dyn. Astron. \textbf{104}, 69–84 (2009a)

\bibitem[\protect\citeauthoryear{Scharz et al.}{2009b}]{SSD09b} Schwarz, R., S\"{u}li, \`{A}., Dvorac, R.: Dynamics of possible Trojan planets in binary systems. Mon. Not. R. Astron. Soc. \textbf{398}, 2085-2090 (2009b)

\bibitem[\protect\citeauthoryear{Sim\'{o}}{1995}]{S78} Sim\'{o}, C.: Relative equilibrium solutions in the four-body problem, Cel. Mech. \textbf{18}, 165-184 (1978)

\bibitem[\protect\citeauthoryear{Sim\'{o} et al.}{1995}]{SGJM95} Sim\'{o}, C., G\'{o}mez, G., Jorba, A., Masdemont, J.: The bicircular model near the triangular libration points of the RTBP, in: From Newton to Chaos (Cortina d'Ampezzo, 1993), NATO Advanced Science Institutes Series B: Physics, vol. 336, Plenum, New York, USA, pp. 343–370 (1995)

\bibitem[\protect\citeauthoryear{Singh \& Vincent}{2015}]{SV15} Singh, J., Vincent, A.E.: Out-of-plane equilibrium points in the photogravitational restricted four-body problem. Astrophys. Space Sci. \textbf{359}, 38 (2015)

\bibitem[\protect\citeauthoryear{Singh \& Vincent}{2016}]{SV16} Singh, J., Vincent, A.E.: Equilibrium points in the restricted four-body problem with radiation pressure. Few-Body Sysytems \textbf{57}, 83-91 (2016)

\bibitem[\protect\citeauthoryear{Soulis et al.}{2008}]{SPB08} Soulis, P.S., Papadakis, K.E., Bountis, T.: Periodic orbits and bifurcations in the Sitnikov four-body problem, Celest. Mech. Dyn. Astron. \textbf{100}, 251-266 (2008)

\bibitem[\protect\citeauthoryear{Szebehely}{1967}]{S67} Szebehely, V.: Theory of orbits. Academic Press, New York (1967)

\bibitem[\protect\citeauthoryear{Van Hamme \& Wilson}{1986}]{VHW86} Van Hamme, W., Wilson, R.E.: The restricted four-body problem and epsilon Aurigae. Astrophys. J. \textbf{306}, L33–L36 (1986)

\bibitem[\protect\citeauthoryear{Wolfram}{2003}]{W03} Wolfram, S.: The Mathematica Book. Fifth Edition, Wolfram Media, Champaign (2003)

\bibitem[\protect\citeauthoryear{Zotos}{2016a}]{Z16a} Zotos, E.E.: Fractal basins of attraction in the planar circular restricted three-body problem with oblateness and radiation pressure. Astrophys. Space Sci. \textbf{361}, article id. 181, 17 pp. (2016a)

\bibitem[\protect\citeauthoryear{Zotos}{2016b}]{Z16b} Zotos, E.E.: Fractal basin boundaries and escape dynamics in a multiwell potential. Nonlin. Dyn. \textbf{85}, 1613-1633 (2016b)

\end{thebibliography}
\end{document}